\documentclass[twocolumn,showpacs,amsmath,amssymb]{revtex4-1}
\usepackage{bm}
\usepackage[T1]{fontenc}
\usepackage{hyperref}
\input{epsf}

\usepackage{graphicx}
\usepackage{epstopdf}
\usepackage{array}
\usepackage{longtable}
\usepackage{rotating,booktabs}
\usepackage{booktabs,threeparttable}
\usepackage{bm}
\usepackage{float}
\usepackage{amsmath}
\usepackage{gensymb}
\usepackage{multirow}
\usepackage{epsfig}
\usepackage{color}

\begin{document}
\title{Hyperfine-structure  constants of the $^{45}\!$Sc II ion and the nuclear quadrupole moment}
\author{Yong-Bo Tang$^{1,*}$, Yu-Shan Zhang$^{1,2}$, and Kai Wang$^{3}$}

\affiliation {$^1$College of Engineering Physics, Shenzhen Technology University, Shenzhen, 518118, China}
\affiliation {$^2$Hebei Key Lab of Optic-electronic Information and Materials, The College of Physics Science and Technology, Hebei University, Baoding 071002, China}
\affiliation {$^3$Department of Physics, Anhui Province Key Laboratory for Control and Applications of Optoelectronic Information Materials, Key Laboratory of Functional Molecular Solids, Ministry of Education, Anhui Normal University, Wuhu, Anhui, 241000, China}
\email{tangyongbo@sztu.edu.cn}
\date{\today}

\begin{abstract}
In this work, we calculate the hyperfine-structure constants of the $^{45}$Sc$^{+}$ ion using a relativistic hybrid approach that combines configuration-interaction and coupled-cluster singles-and-doubles methods. Magnetic-dipole and electric-quadrupole hyperfine-structure constants are determined for the states arising from the $3d4s$, $3d^{2}$, $4s^{2}$, $4s4p$, $3d4p$, $3d5s$, $3d4d$, and $3d5p$ configurations. For most of these states, our magnetic-dipole hyperfine-structure constants agree well with available experimental data and represent a substantial improvement over previous theoretical results. By combining our calculated electric-field gradients with the measured electric-quadrupole hyperfine-structure constants for the $^{3}F_{2,3,4}$, $^{3}P_{1,2}$, and $^{1}G_{4}$ states within the $3d^{2}$ configuration, we derive a nuclear quadrupole moment $Q = - 0.222(5)$ b, which is fully consistent with the value recently obtained from molecular data ( J. P. Dognon and P. Pyykk\"{o}, Phys. Chem. Chem. Phys. 27, 20453 (2025).).
\end{abstract}

\maketitle
\section{Introduction}\label{sec1}
Hyperfine interaction originates from the coupling between the nuclear
electromagnetic multipole moments and the electrons, splitting the fine
structure into the hyperfine structure. Within the first-order perturbation
approximation, the hyperfine structure energy can be expressed as products
of hyperfine-structure (HFS) constants and their corresponding angular-momentum
geometric factors (Clebsch--Gordan coefficients)~\cite{Schwartz}. Accurate
calculations of these constants serve several purposes: for instance,
combining accurate theoretical results with high-precision hyperfine
spectra allows extraction of the nuclear magnetic-dipole, electric-quadrupole,
and even magnetic-octupole moments~\cite{Stone2005ADNDT,Pekka2018MP,Li2021pra,Groote2021pra};
the magnetic-dipole HFS constant is sensitive to the amplitude of the wave
function at the nucleus, so agreement between experiment and theory provides
a stringent test of wave-function quality near the nucleus and, indirectly,
gauges the reliability of the same wave function when used to compute EDM
or PNC amplitudes~\cite{ginges2004violations,safronova2018search}. In
astrophysics, accurate HFS constants are essential for
determining elemental abundances; without them, abundance errors can reach
$2$--$3$ orders of magnitude~\cite{del2017hyperfine,pickering1996measurements}.
Some high-lying atomic states are not readily accessible to experiment, so
theoretically calculated hyperfine-structure constants are
required~\cite{xu2021studies,nave2022hyperfine}.

Scandium (Sc) is the lightest element in the iron group. Its only stable
isotope has nuclear spin $I=7/2$, atomic number $Z=21$, and mass number
$A=45$. Because of its relatively high stellar abundance and rich line
spectrum, accurate knowledge of the HFS constants of Sc
and its ions is essential for astrophysical
research~\cite{thorsbro2018evidence,lawler2019transition}. Given their
importance, the HFS constants of many states in Sc~I and
Sc~II have been systematically
measured~\cite{nave2022hyperfine,xu2021studies,arnesen1982hyperfine,mansour1989laser,
young1988hyperfine,Villemoes1992pra,bieron1997nuclear,bieron1995large,chen1994core,dockery2023hyperfine}.
The present work focuses primarily on the HFS constants
of Sc~II.

For Sc~II, experimental values of the magnetic-dipole ($A$) and
electric-quadrupole ($B$) HFS constants have been reported for
$24$ levels in the $3d4s$, $3d^2$, $4s4p$, and $3d4p$ configurations.
Arnesen first resolved the hyperfine structure of Sc~II by collinear
laser-ion-beam spectroscopy and reported HFS constants
for $3d^2\,{}^1\!D_2$, ${}^3\!P_{1,2}$, and $3d4p\,{}^1\!D_2^{o}$,
${}^3\!D_{1-3}^{o}$ states~\cite{arnesen1982hyperfine}. Young~\textit{et
al.} remeasured the HFS constants $A$ and
$B$ of these levels with collinear fast-ion laser-rf double
resonance~\cite{young1988hyperfine}, added data for the
$3d^2\,{}^1\!G_4$ and $3d4p\,{}^3\!F_{2,3}^{o}$, ${}^3\!P_{1,2}^{o}$,
${}^1\!F_{3}^{o}$ states, and performed multiconfiguration Dirac-Hartree-Fock (MCDF) calculations; noticeable
discrepancies between experiment and theory remained for the triplets.
Subsequently, Mansour~\textit{et al.} improved the experimental precision
by two orders of magnitude and provided HFS constants $A$ and
$B$ for the $3d^2\,{}^3\!F_{2-4}$ states for the first time.
Villemoes~\textit{et al.} extended the measurements to
$3d4p\,{}^3\!F_{4}^{o}$ and $3d4s\,{}^3D_{1-3}$, ${}^1D_2$
states~\cite{Villemoes1992pra}; their MCDF modeling reduced, but did not
remove, the earlier discrepancies. Chen included core-core correlation
and nuclear-spin polarization in multiconfiguration Hartree-Fock and configuration interaction (MCHF-CI)
 calculations and obtained the
HFS constants $A$ and $B$ for the $3d^2\,{}^1\!D_2$, ${}^3\!P_{1,2}$,
${}^1\!G_4$, ${}^3\!F_{2-4}$ states that agree closely with the accurate
experimental values~\cite{chen1994core}.
Xu~\textit{et al.}~\cite{xu2021studies} redetermined the HFS constants
$A$ for $3d4p\,{}^3\!F_{2,3}^{o}$, ${}^1\!D_2^{o}$, ${}^1P_{1}^{o}$
states from Fourier-transform spectra and presented the first data for
the high-lying $3d5s\,{}^3\!D_{1-3}$, $3d4d\,{}^3\!G_3$,
$3d5p\,{}^3\!D_{1,2}^{o}$, ${}^3\!F_3^{o}$, and $3d6s\,{}^3D_2$ levels,
albeit with large uncertainties. Hala and Nave~\cite{nave2022hyperfine}
remeasured the HFS constants of $3d4p\,{}^1\!P_{1}^{o}$,
${}^3\!G_{3}^{o}$ states and added the $4s4p\,{}^3\!P_{1,2}^{o}$ and
$3d4d\,{}^3\!G_{4,5}^{o}$ states, again with moderate precision.
Most recently, Dockery~\textit{et al.} performed collinear laser
spectroscopy on Sc~II and determined the HFS constants $A$ and $B$ for
the $3d4s\,{}^3\!D_{1-3}$ and $3d4p\,{}^3\!F_{2-4}^{o}$,
${}^3\!D_{1-3}^{o}$ states at significantly enhanced
precision~\cite{dockery2023hyperfine}.

In the above reports, the experimental HFS constants \(A\) for most states agree with one another and are precise enough to test theoretical calculations. In contrast, for the HFS constants \(B\) the experimental values for some states differ markedly. Systematic theoretical calculations of HFS constants for Sc~II are still scarce compared with experiments, and published MCDF values~\cite{young1988hyperfine, Villemoes1992pra} differ markedly from measurements: their magnitudes and even the signs are inconsistent. Therefore, a systematic and comprehensive \textit{ab initio} calculation of the hyperfine-structure constants for this ion is needed.

 Recently, Tang developed a relativistic hybrid-method code for accurate calculations of divalent-atomic properties~\cite{Tang2025pra}. The hybrid approach combines the strengths of configuration-interaction (CI) and coupled-cluster (CC) methods, simultaneously treating core-core, core-valence, and valence-valence correlations. It has been used to compute hyperfine-interaction parameters for the low-lying states of \(^{43}\mathrm{Ca}\), \(^{25}\mathrm{Mg}\), \(^{87}\mathrm{Sr}\), and \(^{135,137}\mathrm{Ba}\), confirming that this hybrid approach captures most of the electron-correlation effects and yields accurate theoretical values~\cite{Tang2025pra,Tang2026pra}. In the present work we apply this hybrid method to determine the HFS constants \(A\) and \(B\) of the levels belonging to the \(3d4s\), \(3d^2\), \(4s^2\), \(4s4p\), \(3d4p\), \(3d5s\), \(3d4d\), and \(3d5p\) configurations of Sc~II. Unlike in the regular \(4s^2\rightarrow 4s4p\rightarrow 4s3d\) ladder in neutral Ca, the level structure of Sc~II is complicated by the competition between the \(3d\) and \(4s\) orbitals for the lowest energy, which produces more than a dozen even-parity metastable states from the \(3d^2\) and \(3d4s\) configurations. Accurately calculating the properties of these metastables is more challenging than those for the \(4s^2\) and \(4s4p\) configurations, making Sc~II an excellent test case for relativistic atomic many-body methods.

 The paper is organized as follows. The theoretical framework and definitions of the HFS constants are presented in Sec.~\ref{theory}. Numerical results and comparison with available experimental and theoretical data are reported in Sec.~\ref{results}. The main findings are summarized in Sec.~\ref{conclusions}. Atomic units are used throughout unless otherwise noted.

\section{THEORETICAL EQUATIONS}\label{theory}

\subsection{Method}

In a many-electron atomic system, electrons are conventionally divided into core and valence sets. Consequently, electron-electron correlation is partitioned into core-core, core-valence, and valence-valence components, all of which are essential for accurate calculations of hyperfine interaction properties. In the present work, we treat these three correlation classes with the relativistic CI+CC approach. Within this scheme, we first carry out a self-consistent Dirac-Fock (DF) calculation in a finite basis set to generate the full set of single-particle orbitals; these orbitals are used to construct the model space for subsequent correlation calculations. Next, a CC calculation is performed to construct the so-called correlation potentials that encapsulate core-core and core-valence correlations. Valence-valence correlation is then included via another CI calculation. Finally, the resulting wave functions and energies are employed to evaluate various atomic properties.

For Sc~II, the equation for the effective interaction can be formulated as follows:
 \begin{equation}\label{1}
     \left(\sum_{i = 1}^{2}(H_{\rm DF}+\Sigma_{1})+\frac{1}{r_{12}}+\Sigma_{2}\right)\vert\gamma JM\rangle = E\vert\gamma JM\rangle,
 \end{equation}
 where $H_{\rm DF}$ is the DF Hamiltonian, $\Sigma_{1}$ and $\Sigma_{2}$ are the one- and two-body correlation potentials, respectively, and $\gamma$ represents the quantum numbers other than $J$ and $M$. Here we give the effective equation for the valence electrons only; the core electrons are absorbed into the one- and two-body correlation potentials, and the interaction equations for the many-electron system are detailed in Ref.~\cite{Dzuba1996pra}.

 In the present work, we utilize the coupled-cluster method with single and double excitations (CCSD) to formulate the one-body and two-body correlation potentials. This approach is designated as the CI+CCSD method. To explore the correlation trend, we also employ the second-order many-body perturbation theory (MBPT(2)) and the linear version of the coupled-cluster method with single and double excitations (LCCSD) for the construction of the one-body and two-body correlation potentials. These methods are respectively denoted as CI+MBPT(2) and CI+LCCSD. To account for the remaining higher-order correlation effects, we introduce a scaling parameter $\rho_{\kappa}$, with $\kappa = \ell(\ell+1) - j(j+1) + 1/4$ being the relativistic quantum number related to $j$ and $\ell$, and substitute the one-body correlation potential $\Sigma_1$ by $\rho_{\kappa}\Sigma_1$.
 By adjusting the value of this parameter, the calculated energy can be brought closer to the experimental value. The scheme employing the correlation potential constructed from the LCCSD method is denoted as CI+LCCSDs, while that based on the CCSD method is denoted as CI+CCSDs. Although the CI+LCCSDs and CI+CCSDs methods depend on experimental energies, they remain valuable tools. They can provide high-lying excitation energies with reasonable accuracy. Furthermore, by comparing results obtained with and without the scaling parameter, we can assess how sensitive the property is to higher-order correlation, yielding a rough uncertainty estimate. In the present work, we mainly use these two methods to assess the uncertainty of hyperfine-interaction parameters.

We then compute the transition matrix elements of various operators using the obtained wave functions. In these calculations, we incorporate the random-phase approximation, core Brueckner effects, structural radiation, and normalization corrections of all orders~\cite{Tang2017pra,Blundell1989pra,Dzuba1998jetp}. In addition, we also include two-particle interactions~\cite{Dzuba1998jetp,Safronova1999jpb,Savukov2004pra}. The core Brueckner effects, structural radiation, normalization corrections
and two-particle (TP) interactions are essential for an accurate evaluation of the hyperfine properties.
The method and its underlying formulas are described in detail in a recent work by Tang~\cite{Tang2025pra}.

The relativistic configuration interaction plus a linearized version of coupled-cluster theory, referred to as the RCI+all-order method, was first developed by Safronova \textit{et al.}~\cite{Safronova2009pra}; subsequently, Dzuba independently devised a similar approach~\cite{Dzuba2014pra}. The concept of this hybrid approach can be traced back to the RCI+MBPT combination introduced by Dzuba~\textit{et al.}~\cite{Dzuba1996pra}. The method employed in our work is conceptually similar to theirs; however, we have independently developed our own software package for this hybrid method, and consequently there remain a few differences in the handling details compared with their implementations.

\subsection{Hyperfine-structure constant}
Typically, hyperfine splitting is orders of magnitude smaller than fine-structure splitting, allowing the hyperfine interaction to be treated as a perturbation. Retaining only the first-order correction, its energy is parameterized as:
\begin{small}
\begin{align}
\Delta{E_F^{(1)}}=&\frac{A}{2}K+\frac{B}{2}\frac{3K(K+1)-4I(I+1)J(J+1)}{2I(2I-1)2J(2J-1)},
\end{align}
\end{small}
where $K=F(F+1)-I(I+1)-J(J+1)$,  $A$ and $B$ are the magnetic dipole and the electric quadrupole HFS constant, which are defined as~\cite{Schwartz}:
\begin{small}
\begin{align}
A=\frac{\mu}{I}\frac{\langle\gamma J\|T^{(1)}\|\gamma J\rangle}{\sqrt{J(J+1)(2J+1)}},
\end{align}
\end{small}
and
\begin{small}
\begin{align}
B=2Q\bigg[\frac{2J(2J-1)}{(2J+1)(2J+2)(2J+3)}\bigg]^{1/2}\langle\gamma J\|T^{(2)}\|\gamma J\rangle,
\end{align}
\end{small}
respectively, and $T^{(k)}=\sum_{i}{t^{(k)}(\textbf{r}_{i})}$. The single-particle reduced matrix elements of the operators $t^{(1)}$ and $t^{(2)}$ are given by:
\begin{align}
\langle\kappa_{a}\|t^{(1)}\|\kappa_{b}\rangle
={}&-(\kappa_{a}+\kappa_{b})\,
   \langle-\kappa_{a}\|C^{(1)}\|\kappa_{b}\rangle\notag\\
&\times\int_{0}^{\infty}
   \frac{f_{a}(r)g_{b}(r)+f_{b}(r)g_{a}(r)}{r^{2}}\,
   F^{(1)}(r)\,dr,
\end{align}
and
\begin{align}
\langle\kappa_{a}\|t^{(2)}\|\kappa_{b}\rangle
={}&-\langle\kappa_{a}\|C^{(2)}\|\kappa_{b}\rangle\notag\\
&\times\int_{0}^{\infty}
   \frac{f_{a}(r)f_{b}(r)+g_{a}(r)g_{b}(r)}{r^{3}}\,
   F^{(2)}(r)\,dr.
\end{align}
Here, the nuclear distribution function $F^{(k)}(r)$ is defined as:
\begin{align}
	F^{(k)}(r) =\begin{cases}
		(\frac{r}{R_{N}})^{2k+1},&r\leq{R_{N}}\\
		1,&r>R_{N}
	\end{cases},
\end{align}
where $R_{N}=\sqrt{5/3}\langle{r^{2}}\rangle^{1/2}$ is the radius of the sphere, and $\langle{r^{2}}\rangle^{1/2}$ is the charge root-mean-square radius of the nucleus. In the present work, the nucleus is modeled as a uniformly magnetized and charged sphere.

\subsection{Computation details}
 In Dirac-Fock calculations, the nuclear charge is described by a Fermi distribution, and the one-electron orbitals are expanded in a finite even-tempered Gaussian basis:
 \begin{align}
 G_{i\kappa}(r) =  \mathcal{N}_i r^{n_\kappa} e^{-\alpha_i r^2},
 \end{align}
 where $\mathcal{N}_i$ is the normalization constant, $n_\kappa=\ell+1$, and $\alpha_i=\alpha\beta^{i-1}$.

\begin{table}
\caption{The parameters of the Gauss basis set, along with the basis spaces employed in the CC and CI calculations.}
\label{tab1}
\begin{ruledtabular}
\begin{tabular}{ccccccccccc}
 &$s$&$p$&$d$&$f$&$g$&$h$&$i$\\
\hline
$\alpha$&0.00058&0.00057&0.00056&0.00060&0.086&0.096&0.096\\
$\beta$&1.95&1.94&1.96&1.99&2.0&2.2&2.2\\
N    &45&40&40&35&30&25&15\\
N$_{\rm core}$    &1-3&1-2&0&0&0&0&0\\
N$_{\rm valence}$    &4-8&3-7&1-5&1-3&0&0&0\\
N$_{\rm virtual}$  &4-30&3-28&1-26&1-22&1-15&1-13&1-12\\
N$_{\rm CI}$   &4-22&3-21&1-21&1-19&1-15&1-13&0\\
\end{tabular}
\end{ruledtabular}\label{tab2}
\end{table}

Table~\ref{tab1} lists the exponents $\alpha$, $\beta$, and the basis size $N$ for each symmetry. $N_{\text{core}}$, $N_{\text{val}}$, and $N_{\text{virt}}$ denote the numbers of orbitals treated as core, valence, and virtual in coupled-cluster (CC) calculations; $N_{\text{CI}}$ gives the number of configurations retained for CI expansions. The highest partial wave included is $\ell_{\text{max}}=6$ for CC and $\ell_{\text{max}}=5$ for CI calculations; in second-order many-body perturbation theory (MBPT(2)) the complete basis is used. The sizes of the Gaussian basis set, virtual orbital space, and configurations quoted in Table~\ref{tab1} are fully adequate: enlarging any of these spaces changes the final energies and hyperfine-structure constants by less than one unit in the last quoted digit.

The scaling parameters for the CI+LCCSDs method are $\rho_{-1}=0.988$, $\rho_{1}=0.969$, $\rho_{-2}=0.969$, $\rho_{2}=0.933$, $\rho_{-3}=0.933$, and $\rho_{\text{other}}=1.0$; those for the CI+CCSDs method are $\rho_{-1}=0.982$, $\rho_{1}=1.0$, $\rho_{-2}=1.0$, $\rho_{2}=0.992$, $\rho_{-3}=0.992$, and $\rho_{\text{other}}=1.0$. The values of these parameters are determined through a two-step process. First, the rescaling parameter $\rho_{\kappa}$ is adjusted to reproduce the experimental energy of the lowest state for each angular quantum number $\kappa$ of the monovalent atomic system Sc$^{2+}$, a procedure similar to the Dirac-Fock plus core polarization (DFCP) method~\cite{Tang2013pra}. Subsequently, the value of $\rho_{\kappa}$ is further refined to bring the energy calculated using either the CI+LCCSDs or CI+CCSDs method into closer agreement with the experimental value for Sc$^+$.

\section{Results and Discussion}\label{results}
\subsection{Ionization energies}

Table~\ref{tab2} lists  the ionization energies  of 58 states arising from the $3d4s$, $3d^{2}$, $4s^{2}$, $4s4p$, $3d4p$, $3d5s$, $3d4d$, and $3d5p$ configurations of the Sc II ion. These energies are obtained by using CI+MBPT(2), CI+LCCSD, CI+CCSD, CI+LCCSDs and CI+CCSDs methods, respectively.  We also compare our calculated results with the experimental values from the National Institute of Standards and Technology (NIST)~\cite{NIST_ASD}. For convenience, the CI+MBPT(2), CI+LCCSD, CI+CCSD, CI+LCCSDs and CI+CCSDs methods are denoted as $\rm M_{1}$, $\rm M_{2}$, $\rm M_{3}$, $\rm M_{4}$, and $\rm M_{5}$ respectively. The symbol $\Delta_{n}$ represents the difference between the theoretical results obtained by the $\rm M_{n}$ method and the experimental values, i.e., $\Delta_{n}=E_{\rm M_{n}}-E_{\rm NIST}$. As seen, CI+MBPT(2) method yields the largest mean deviation, 2380 cm$^{-1}$; CI+LCCSD reduces this to 905 cm$^{-1}$; and CI+CCSD further lowers it to 110 cm$^{-1}$, markedly outperforming the CI+MBPT(2) and CI+LCCSD methods. The mean deviations for CI+LCCSDs and CI+CCSDs methods are 200 cm$^{-1}$ and 80 cm$^{-1}$, respectively.
For low-lying states, the CI+LCCSDs and CI+CCSDs methods yield better results than the CI+CCSD method. However, across all states, the CI+CCSD method remains relatively stable, with deviations of approximately 200 cm$^{-1}$ for each state. This is understandable given that our fitting parameters are typically determined by adjusting against the first state of each $J$ level as the reference. The discrepancy between our CI+CCSD results and the experimental values arises from high-order effects not taken into account in the present work.

\begin{table*}[!]
    \setlength\tabcolsep{2pt}
    \addtolength{\tabcolsep}{3pt}
\caption[]{Ionization energies (in cm$^{-1}$) for Sc II levels in the $3d4s$, $3d^{2}$, $4s^{2}$, $4s4p$, $3d4p$, $3d5s$, $3d4d$, and $3d5p$ configurations. }
\begin{ruledtabular}
 \begin{tabular}{lcccccccccccccccccc}
Conf.      &    Term     & $E_{\rm M_{1}}$  &$E_{\rm M_{2}}$ &$E_{\rm M_{3}}$&$E_{\rm M_{4}}$ &$E_{\rm M_{5}}$ & $E_{\rm NIST}$   &  $\Delta_{1}$   & $\Delta_{2}$ &$\Delta_{3}$ & $\Delta_{4}$ &$\Delta_{5}$ \\
\hline
3d4s & $^3\!D_1$ & 305461    & 303908  & 303086  & 302921  & 302901  & 302914  &--2546  &--993  &--171  &--6  & 14 \\
3d4s & $^3\!D_2$ & 305390    & 303837  & 303015  & 302853  & 302830  & 302847  &--2543  &--990  &--168  &--6  & 16 \\
3d4s & $^3\!D_3$ & 305275    & 303724  & 302903  & 302746  & 302719  & 302737  &--2538  &--987  &--167  &--9  & 18 \\
3d4s & $^1\!D_2$ & 303048    & 301399  & 300570  & 300361  & 300390  & 300374  &--2675  &--1025  &--196  &--13  & 17 \\
$3d^{2} $& $^3\!F_2$ & 301712    & 299755  & 298353  & 298128  & 298169  & 298112  &--3600  &--1643  &--242  &--17  & 58 \\
$3d^{2} $& $^3\!F_3$ & 301625    & 299666  & 298266  & 298044  & 298082  & 298031  &--3595  &--1635  &--235  &--13  & 51 \\
$3d^{2} $& $^3\!F_4$ & 301510    & 299564  & 298163  & 297948  & 298163  & 297927  &--3584  &--1637  &--237  &--21  & 237 \\
$3d^{2} $& $^1\!D_2$ & 294987   & 293557  & 292200  & 292021  & 292020  & 291970  &--3017  &--1587  &--230  &--51  & 50 \\
$4s^{2} $ & $^1\!S_0$ & 293917    & 292455  & 291569  & 291760  & 291388  & 291178  &--2739  &--1277  &--391  &--582  & 210 \\
$3d^{2} $& $^3\!P_0$ & 292822   & 292011  & 291057  & 290880  & 290879  & 290840  &--1981  &--1170  &--216  &--40  & 38 \\
$3d^{2} $& $^3\!P_1$ & 293884   & 292414  & 291019  & 290842  & 290841  & 290813  &--3071  &--1601  &--206  &--29  & 28 \\
$3d^{2} $& $^3\!P_2$ & 293824   & 292357  & 290962  & 290789  & 290785  & 290760  &--3064  &--1597  &--202  &--29  & 25 \\
$3d^{2} $& $^1\!G_4$ & 291930   & 290205  & 288840  & 288652  & 288840  & 288653  &--3277  &--1552  &--187  & 1  & 187 \\
$3d^{2} $& $^1\!S_0$ & 279126    & 278143  & 277048  & 276846  & 276892  & 276959  &--2167  &--1183  &--88  & 113  & 68 \\
3d4p & $^1\!D^{o}_2$ & 279253    & 277804  & 277046  & 276777  & 276939  & 276833  &--2420  &--971  &--213  & 57  & 106 \\
3d4p & $^3\!F^{o}_2$ & 277742    & 276266  & 275539  & 275246  & 275433  & 275471  &--2271  &--795  &--68  & 224  & 38 \\
3d4p & $^3\!F^{o}_3$ & 277580    & 276101  & 275376  & 275085  & 275270  & 275312  &--2268  &--789  &--64  & 227  & 42 \\
3d4p & $^3\!F^{o}_4$ & 277334   & 275858  & 275134  & 274847  & 275134  & 275073  &--2261  &--785  &--61  & 226  & 61 \\
3d4p & $^3\!D^{o}_1$ & 277481   & 275930  & 275188  & 274883  & 275078  & 274997  &--2485  &--933  &--191  & 114  & 81\\
3d4p & $^3\!D^{o}_2$ & 277374    & 275826  & 275084  & 274782  & 274974  & 274893  &--2481  &--932  &--191  & 111  & 81 \\
3d4p & $^3\!D^{o}_3$ & 277229    & 275681  & 274941  & 274642  & 274831  & 274753  &--2475  &--928  &--188  & 111  & 78 \\
3d4p & $^3\!P^{o}_0$ & 275278    & 273916  & 273227  & 272961  & 273122  & 273178  &--2100  &--738  &--49  & 218  & 56 \\
3d4p & $^3\!P^{o}_1$ & 275275    & 273909  & 273220  & 272954  & 273115  & 273172  &--2103  &--737  &--48  & 219  & 58 \\
3d4p & $^3\!P^{o}_2$ & 275194    & 273830  & 273138  & 272875  & 273033  & 273091  &--2103  &--739  &--48  & 215  & 57\\
3d4p & $^1\!P^{o}_1$ & 274280    & 272785  & 272124  & 271850  & 272018  & 272099  &--2182  &--686  &--26  & 248  & 80 \\
3d4p & $^1\!F^{o}_3$ & 272914    & 271263  & 270566  & 270215  & 270454  & 270564  &--2349  &--698  &--1  & 349  & 111 \\
4s4p & $^3\!P^{o}_0$ & 265057   & 264431  & 264104  & 264169  & 264005  & 263912  &--1145  &--519  &--191  &--257  & 93 \\
4s4p & $^3\!P^{o}_1$ & 264940    & 264324  & 263998  & 264064  & 263900  & 263799  &--1141  &--524  &--199  &--265  & 100 \\
4s4p & $^3\!P^{o}_2$ & 264699    & 264082  & 263759  & 263828  & 263660  & 263569  &--1130  &--513  &--190  &--259  & 91 \\
4s4p & $^1\!P^{o}_1$ & 248716    & 247928  & 247505  & 247611  & 247401  & 247199  &--1517  &--729  &--306  &--412  & 201 \\
3d5s & $^3\!D_1$ & 247718    & 246154  & 245484  & 245167  & 245356  & 245363  &--2356  &--792  &--122  & 196  & 7 \\
3d5s & $^3\!D_2$ & 247653   & 246090  & 245421  & 245106  & 245292  & 245300  &--2353  &--790  &--121  & 194  & 8 \\
3d5s & $^3\!D_3$ & 247518    & 245956  & 245287  & 244978  & 245159  & 245171  &--2347  &--785  &--117  & 193  & 11 \\
3d5s & $^1\!D_2$ & 247004   & 245445  & 244783  & 244470  & 244656  & 244662  &--2342  &--783  &--121  & 193  & 6 \\
3d4d & $^1\!F_3$ & 245718   & 244110  & 243371  & 243060  & 243252  & 243386  &--2332  &--724  & 15  & 326  & 134 \\
3d4d & $^3\!D_1$ & 245371   & 243764  & 243027  & 242713  & 242908  & 243039  &--2332  &--724  & 12  & 326  & 131 \\
3d4d & $^3\!D_2$ & 245315    & 243710  & 242974  & 242662  & 242855  & 242985  &--2330  &--725  & 11  & 323  & 130 \\
3d4d & $^3\!D_3$ & 245239    & 243633  & 242898  & 242589  & 242779  & 242913  &--2327  &--721  & 15  & 323  & 133 \\
3d4d & $^3\!G_3$ & 244956    & 243357  & 242619  & 242306  & 242500  & 242647  &--2309  &--709  & 28  & 342  & 147 \\
3d4d & $^3\!G_4$ & 244872    & 243277  & 242540  & 242229  & 242540  & 242566  &--2306  &--711  & 26  & 337  & 26 \\
3d4d & $^1\!P_1$ & 244819    & 243234  & 242494  & 242188  & 242376  & 242514  &--2305  &--720  & 20  & 326  & 138 \\
3d4d & $^3\!G_5$ & 244759   & 243161  & 242424  & 242118  & 242424  & 242457  &--2302  &--704  & 33  & 339  & 33 \\
3d4d & $^3\!S_1$ & 244139    & 242558  & 241825  & 241518  & 241707  & 241843  &--2296  &--715  & 18  & 325  & 136 \\
3d4d & $^3\!F_2$ & 241879    & 240304  & 239520  & 239232  & 239399  & 239540  &--2339  &--764  & 20  & 308  & 141 \\
3d4d & $^3\!F_3$ & 241805    & 240231  & 239447  & 239162  & 239325  & 239469  &--2336  &--762  & 23  & 307  & 144 \\
3d4d & $^3\!F_4$ & 241718    & 240148  & 239365  & 239083  & 239365  & 239386  &--2332  &--762  & 21  & 303  & 21 \\
3d4d & $^1\!D_2$ & 240867    & 239359  & 238579  & 238295  & 238458  & 238548  &--2320  &--811  &--31  & 252  & 90 \\
3d4d & $^3\!P_0$ & 240541   & 239071  & 238291  & 238019  & 238172  & 238299  &--2243  &--772  & 8  & 280  & 127\\
3d4d & $^3\!P_1$ & 240510    & 239039  & 238258  & 237988  & 238140  & 238268  &--2242  &--771  & 9  & 280  & 128 \\
3d4d & $^3\!P_2$ & 240452    & 238983  & 238204  & 237934  & 238085  & 238209  &--2243  &--775  & 5  & 274  & 124 \\
3d4d & $^1\!G_4$ & 240065    & 238501  & 237718  & 237432  & 237718  & 237678  &--2386  &--822  &--39  & 247  & 39 \\
3d5p & $^1\!D^{o}_2$ & 239239    & 237709  & 237044  & 236710  & 236933  & 236866  &--2373  &--843  &--178  & 156  & 66 \\
3d5p & $^3\!D^{o}_1$ & 238915    & 237363  & 236702  & 236360  & 236590  & 236525  &--2390  &--839  &--177  & 165  & 65\\
3d5p & $^3\!F^{o}_2$ & 238780    & 237237  & 236579  & 236239  & 236468  & 236455  &--2325  &--782  &--124  & 216  & 13 \\
3d5p & $^3\!D^{o}_2$ & 238830    & 237280  & 236620  & 236279  & 236509  & 236422  &--2409  &--859  &--198  & 142  & 87 \\
3d5p & $^3\!F^{o}_3$ & 238685   & 237145  & 236487  & 236149  & 236376  & 236351  &--2334  &--794  &--136  & 202  & 25 \\
3d5p & $^3\!D^{o}_3$ & 238724   & 237175  & 236515  & 236178  & 236404  & 236331  &--2393  &--845  &--184  & 152  & 73 \\
3d5p & $^3\!F^{o}_4$ & 238535   & 236996  & 236339  & 236004  & 236339  & 236195  &--2339  &--801  &--144  & 191  & 144\\

\end{tabular}\label{tab2}
\end{ruledtabular}
\end{table*}

\subsection{Magnetic dipole hyperfine-structure constants}
Table~\ref{tab3} presents the HFS constants $A$ calculated using five methods ($\rm M_{1}$--$\rm M_{5}$, as defined in Table~\ref{tab2}). Overall, CI+MBPT(2) differs markedly from the four CI+CC variants: for most states, deviations range from 3-20\%, highlighting the importance of electron correlation beyond second order. For the four states $3d^{2} \,^{3}\!F_{4}$, $3d^{2} \,^{3}\!P_{2}$, $3d4d \,^{3}\!D_{3}$, and $3d4d \,^{3}\!S_{1}$, discrepancies reach 30\% or more, indicating exceptional sensitivity to electron correlation. The HFS constants of these states are either very small in magnitude or opposite in sign to those of other states. The four CI+CC approaches generally agree within 2\%. Previous studies~\cite{Tang2025pra,Tang2026pra} showed that HFS constants $A$ calculated with CI+CC methods deviate from experimental values by approximately 2\%. Accordingly, we adopt the CI+LCCSD values as our recommended results (denoted by $A_{\rm Rec.}$), with uncertainties are taken as the larger of: (i) the maximum deviation between CI+LCCSD and the other three CI+CC results; (ii) 2\% of the recommended value.

\begin{table}[!]
\caption[]{HFS constants $A$ (in MHz) of $^{45}\!$Sc II. The magnetic dipole moment is $\mu = 4.7559\,\mu_{\rm N}$, where $\mu_{\rm N}$ is the nuclear magneton. The recommended values are denoted by $A_{\rm Rec.}$, with uncertainties given in parentheses. }
\begin{ruledtabular}
 \begin{tabular}{lcccccccccccccccc}
Conf. & 	Term & $A_{\rm M1}$&$A_{\rm M2}$&$A_{\rm M3}$&$A_{\rm M4}$&$A_{\rm M5}$& $A_{\rm Rec.}$\\
\hline
3d4s   & $ ^3\!D_1 $  & --527    & --469  & --462& --473  & --470  & --469(10)\\
       & $ ^3\!D_2 $  & 508   & 507  & 503  &504&503& 507(11)\\
       & $ ^3\!D_3 $  & 669   & 652  & 647  &650&648& 652(12)\\
3d4s   & $ ^1\!D_2 $  & 112  & 126  & 127 &129&128& 126(3)\\
$3d^2$ & $ ^3\!F_2 $  & 299    & 286  & 285  &285&286& 286(6)\\
       & $ ^3\!F_3 $  & 97   & 113  & 114  &112&112& 113(3)\\
       & $ ^3\!F_4 $  & 9.1   & 41  & 41  &40&39&41(2)\\
$3d^2$ & $ ^1\!D_2 $  & 150    & 152  & 151  &151&150& 152(3)\\
$3d^2$ & $ ^3\!P_1 $  & --152    & --114  & --113  &--114&--113& --114(3)\\
       & $ ^3\!P_2 $  & --88    & --35.8  & --36.3 &--36.2&--36.4& --36(2)\\
$3d^2$ & $ ^1\!G_4 $  & 120    & 137  & 136  &135&136& 137(3)\\
3d4p   & $ ^1\!D^{o}_2 $  & 209   & 213  & 212  &212&213& 213(4)\\
3d4p   & $ ^3\!F^{o}_2 $  & 376    & 369  & 366  &366&366& 369(7)\\
       & $ ^3\!F^{o}_3 $  & 202    & 204  & 201  &200&201& 204(4)\\
       & $ ^3\!F^{o}_4 $  & 96  & 100  & 99  &97&98&100(3)\\
3d4p   & $ ^3\!D^{o}_1 $  & 299   & 304  & 304  & 302&303&304(6)\\
       & $ ^3\!D^{o}_2 $  & 114    & 129  & 128  &128&129&129(3)\\
       & $ ^3\!D^{o}_3 $  & 88.9  & 100  & 99  &97&99&100(3)\\
3d4p   & $ ^3\!P^{o}_1 $  & 209    & 259  & 250  &257&250&259(9)\\
       & $ ^3\!P^{o}_2 $  & 63.9    &103  & 101  &102&101&103(2)\\
3d4p   & $ ^1\!P^{o}_1 $  & 128   & 147  & 145 &142&145&147(5)\\
3d4p   & $	^1\!F^{o}_3 $  & 181   & 187  & 186  & 184&186&187(4)\\
4s4p   & $ ^3\!P^{o}_1 $  & 1444    & 1354  & 1337  &1326&1330&1354(28)\\
       & $ ^3\!P^{o}_2 $  & 1205   & 1135  & 1123  & 1115&1118&1135(30)\\
4s4p   & $ ^1\!P^{o}_1 $  & 208   & 199  & 198  &198&198&199(4)\\
3d5s   & $ ^3\!D_1 $  & 180   & 198  & 197  &194&197&198(4)\\
       & $ ^3\!D_2 $  & 327    & 335  & 333  &330&333&335(7)\\
       & $ ^3\!D_3 $  & 222    & 233  & 231  & 230&230&233(5)\\
3d5s   & $ ^1\!D_2 $  & 95   & 112  & 112  & 115&112&112(3)\\
3d4d   & $ ^1\!F_3 $  & 105    & 111  & 111  &110&111&111(3)\\
3d4d   & $ ^3\!D_1 $  & 221    & 204  & 204  &202&205&204(4)\\
       & $ ^3\!D_2 $  & 52.1    & 63.3  & 62.9  & 62.8&62.9&63(2)\\
       & $ ^3\!D_3 $  & 16.1    & 36.1  & 35.5  & 35.3&35.4&36(2)\\
3d4d   & $ ^3\!G_3 $  & 188    & 185  & 185  &183&185&185(4)\\
       & $ ^3\!G_4 $  & 95.5  & 104  & 103  & 102&103&104(2)\\
       & $ ^3\!G_5 $  & 36.8    & 49.6  & 49.4  & 48.4&49.3&49(2)\\
3d4d   & $ ^1\!P_1 $  & 89   & 92  & 92  &92&92&92(1)\\
3d4d   & $ ^3\!S_1 $  & --146   & --98  & --98  & --95&--97&--98(3)\\
3d4d   & $ ^3\!F_2 $  & 206   & 198  & 198  & 197&198&198(4)\\
       & $ ^3\!F_3 $  &79& 90     & 89  &88&89&90(2)\\
       & $ ^3\!F_4 $  & 28.5    & 45.7  & 45.1  & 44.5&45.2&45(2)\\
3d4d   & $ ^1\!D_2 $  & 98.2   & 104  & 104  & 102&104&104(2)\\
3d4d   & $ ^3\!P_1 $  & --75.1    & --52.0  & --52.2  &--51.5&--52.1&--52(2)\\
3d4d   & $ ^1\!G_4 $  & 98  & 104  & 103  & 103&104&104(2)\\
3d5p   & $ ^1\!D^{o}_2 $  & 212    & 213  & 213  &210&213&213(5)\\
3d5p   & $ ^3\!D^{o}_1 $  & 339    & 339  & 339  & 335&337&339(7)\\
       & $ ^3\!D^{o}_2 $  & 133    & 152  & 155  & 146&156&152(6)\\
       & $ ^3\!D^{o}_3 $  & 68.8    & 87  & 89  & 83&89&87(5)\\
3d5p   & $ ^3\!F^{o}_2 $  & 241  & 242  & 236  & 247&237&242(5)\\
       & $ ^3\!F^{o}_3 $  & 137  & 144  & 141  & 147&143&144(3)\\
       & $ ^3\!F^{o}_4 $  & 56.5    &70  & 70  &68&69&70(2)\\
\end{tabular}\label{tab3}
\end{ruledtabular}
\end{table}

\begin{table}[!]
\caption[]{Comparison of HFS constants $A$ (in MHz)  for $^{45}\!$Sc~II.}
\begin{ruledtabular}
 \begin{tabular}{lcccccccccccccccc}
Conf.                   	&		Term &	$A_{\rm Present}$&	$A_{\rm Expt.}$	&$A_{\rm Ther.}$	\\
\hline
3d4s   &$^3\!D_1$&--469(10)    &--479.51(2)(7)~\cite{dockery2023hyperfine}     &--473.3~\cite{Villemoes1992pra}         \\
	   &         &             &--480(2)~\cite{Villemoes1992pra}           &--479~\cite{Ruczkowski2014jsqrt}	        \\
       &$^3\!D_2$&507(11)      &  507.53(3)(7)~\cite{dockery2023hyperfine}     &  518.8~\cite{Villemoes1992pra} 	    \\
	   &         &             &  510(1)~\cite{Villemoes1992pra}           &  507~\cite{Ruczkowski2014jsqrt}           \\
       &$^3\!D_3$&652(12)       &  656.73(2)(9)~\cite{dockery2023hyperfine}     &  608.5~\cite{Villemoes1992pra}         \\
	   &         &             &  654.8(6)~\cite{Villemoes1992pra}         &  660~\cite{Ruczkowski2014jsqrt}       	\\
3d4s   &$^1\!D_2$&126(3)  &  128.2(8)~\cite{Villemoes1992pra}         &  146.8~\cite{Villemoes1992pra}      	\\
	   &         &             &	                   &  134~\cite{Ruczkowski2014jsqrt}           \\	
$3d^2$ &$^3\!F_2$&286(6)&290.67(4)~\cite{mansour1989laser}          &  277.6~\cite{Villemoes1992pra}       	\\
	   &         &             &	                   &  285~\cite{Ruczkowski2014jsqrt}           \\  	
       &$^3\!F_3$&113(3)  &113.672(6)~\cite{mansour1989laser}         &  137.5~\cite{Villemoes1992pra}     	\\
	   &         &             &	                   &  118~\cite{Ruczkowski2014jsqrt}           \\	
       &$^3\!F_4$&41(2)   &38.357(4)~\cite{mansour1989laser}          &  60.9~\cite{Villemoes1992pra}       	\\
	   &         &             &	                   &  46~\cite{Ruczkowski2014jsqrt}            \\ 	
$3d^2$ &$^1\!D_2$&152(3)  &149.9(3)~\cite{young1988hyperfine}           &  146.0~\cite{Villemoes1992pra}      	\\
	   &         &             &	149.361(4)~\cite{mansour1989laser}     &  146.6~\cite{young1988hyperfine}       	\\
	   &         &             &	          &   146~\cite{Ruczkowski2014jsqrt}      	\\
$3d^2$ &$^3\!P_1$&--114(3)&--107.501(4)~\cite{mansour1989laser}       &--63.6~\cite{Villemoes1992pra}       	\\
	   &         &             &--108.1(4)~\cite{Villemoes1992pra}         &--1.8~\cite{young1988hyperfine}       	\\
	   &         &             &          &--110~\cite{Ruczkowski2014jsqrt}       	\\
       &$^3\!P_2$&--36(2) &--27.2(4)~\cite{Villemoes1992pra}          &--0.03~\cite{Villemoes1992pra}       	\\
	   &         &             & --27.732(4)~\cite{mansour1989laser}       &85.9~\cite{young1988hyperfine}          	\\
	   &         &             & --27.9(4)~\cite{young1988hyperfine}         &--33~\cite{Ruczkowski2014jsqrt}      	    \\
$3d^2$ &$^1\!G_4$&137(3)  &135.23(4)~\cite{mansour1989laser}          &154.6~\cite{Villemoes1992pra}           \\
	   &         &             &135.2(1.6)~\cite{young1988hyperfine}         &143.2~\cite{young1988hyperfine}      	    \\
	   &         &             &	                   &140~\cite{Ruczkowski2014jsqrt}       	    \\
3d4p&$^1\!D^{o}_2$&213(4)  &215.7(8)~\cite{young1988hyperfine}           &202.5~\cite{Villemoes1992pra}       	\\
	   &         &             &215.7(8)~\cite{young1988hyperfine}           &202.4~\cite{young1988hyperfine}       	\\
	   &         &             &	                   &220~\cite{Ruczkowski2014jsqrt}             \\    	
3d4p   &$^3\!F^{o}_2$&369(7)&367.94(3)(5)~\cite{dockery2023hyperfine}       &425.3~\cite{Villemoes1992pra}       	\\
	   &         &             &366.8(3)~\cite{young1988hyperfine}           &309.2~\cite{young1988hyperfine}       	\\
	   &         &             &	                   &363~\cite{Ruczkowski2014jsqrt}             \\   	
       &$^3\!F^{o}_3$&204(4)  &205.61(3)(3)~\cite{dockery2023hyperfine}       &193.7~\cite{Villemoes1992pra}          	\\
	   &         &             &205.4(12)~\cite{young1988hyperfine}          &193.8~\cite{young1988hyperfine}        	\\
	   &         &             &	                   &199~\cite{Ruczkowski2014jsqrt}      	    \\
       &$^3\!F^{o}_4$&100(3)&102.23(3)(1)~\cite{dockery2023hyperfine}         &115.6~\cite{Villemoes1992pra}      	    \\
	   &         &             &102.3(2)~\cite{young1988hyperfine}           &96~\cite{Ruczkowski2014jsqrt}    	        \\
3d4p   &$^3\!D^{o}_1$&304(6)  &304.98(3)(4)~\cite{dockery2023hyperfine}       &267.7~\cite{Villemoes1992pra}       	\\
	   &         &             &304.7(4)~\cite{young1988hyperfine}           &278.3~\cite{young1988hyperfine}       	\\
	   &         &             &307(2)~\cite{Villemoes1992pra}             &304~\cite{Ruczkowski2014jsqrt}       	    \\
       &$^3\!D^{o}_2$&129(3) &124.95(3)(2)~\cite{dockery2023hyperfine}       &149.0~\cite{Villemoes1992pra}     	    \\
	   &         &             & 125.3(2)~\cite{young1988hyperfine}          &165.5~\cite{young1988hyperfine}       	\\
	   &         &             & 125.7(3)~\cite{Villemoes1992pra}          &126~\cite{Ruczkowski2014jsqrt}       	    \\
       &$^3\!D^{o}_3$&100(3)  &99.64(3)(1)~\cite{dockery2023hyperfine}        &128.3~\cite{Villemoes1992pra}     	    \\
	   &         &             & 101.8(6)~\cite{Villemoes1992pra}         &134.4~\cite{young1988hyperfine}     	    \\
	   &         &             &              &103~\cite{Ruczkowski2014jsqrt}             \\	
3d4p&$^3\!P^{o}_1$&259(9) &258(2)~\cite{Villemoes1992pra}             &148.0~\cite{Villemoes1992pra}      	    \\
	   &         &             & 255.0(4)~\cite{young1988hyperfine}          &167.1~\cite{young1988hyperfine}       	\\
	   &         &             &              &255~\cite{Ruczkowski2014jsqrt}             \\	
       &$^3\!P^{o}_2$&103(2) &105.6(5)~\cite{Villemoes1992pra}           &87.2~\cite{Villemoes1992pra}       	    \\
	   &         &             & 106.2(2)~\cite{young1988hyperfine}          &96.6~\cite{young1988hyperfine}       	\\
	   &         &             &             &98~\cite{Ruczkowski2014jsqrt}       \\	
3d4p&$^1\!P^{o}_1$&147(5) &189(60)~\cite{nave2022hyperfine}            &86~\cite{Ruczkowski2014jsqrt}     	\\
3d4p&$^1\!F^{o}_3$&187(4)  &193.1(8)~\cite{Villemoes1992pra}           &184.9~\cite{Villemoes1992pra}    	\\
	   &          &            &190.6                  &199~\cite{Ruczkowski2014jsqrt}      \\	
4s4p   &$^3\!P^{o}_1$&1354(28)&1343(10)~\cite{nave2022hyperfine}  &1008~\cite{Ruczkowski2014jsqrt}      	\\
       &$^3\!P^{o}_2$&1135(30)&1082(20)~\cite{nave2022hyperfine}  & 841~\cite{Ruczkowski2014jsqrt}      	\\
\end{tabular}\label{tab4}
\end{ruledtabular}
\end{table}

Table~\ref{tab4} compares the available experimental measurements with theoretical calculations. HFS constants $A$ have been measured for $24$ states~\cite{nave2022hyperfine,xu2021studies,arnesen1982hyperfine,mansour1989laser,
young1988hyperfine,Villemoes1992pra,bieron1997nuclear,bieron1995large,chen1994core,dockery2023hyperfine}, but only the high-precision experimental values are listed in the Table~\ref{tab4}. With the exception of the $3d4p\;{}^{1}P_{1}$ state, all experimental data are sufficiently accurate to serve as a stringent test of relativistic atomic many-body methods. Overall, our recommended values agree with the experimental measurements to within about 2\% for the vast majority of states. The exceptions are the $3d^2$ $^3\!F_4$  and $3d^2$ $^3\!P_2$ levels, whose HFS constants $A$ are markedly smaller than those of other states and deviate substantially from the measured values. As shown in Table~\ref{tab3}, these two states are highly sensitive to higher-order electron-correlation effects: the CI+MBPT and CI+CC results differ by more than 50\%. Accurate theoretical predictions of their the hyperfine properties of these two states will likely require the inclusion of triple and even quadruple excitations. Compared with earlier theoretical work~\cite{Villemoes1992pra,young1988hyperfine}, the agreement between our results and experimental values has been significantly improved, which benefits from our adoption of a more complete treatment of electron correlation effects. We also compare with the semi-empirical results of Ref.~\cite{Ruczkowski2014jsqrt}, where a large set of experimental measurements was used as reference data to fit the wave functions and subsequently predict the HFS constants of highly excited states. A comparison of Tables~\ref{tab3} and ~\ref{tab4} reveals that for the vast majority of states, the deviations of both CI+LCCSD and CI+CCSD results from experimental measurements are within 2\%. This once again demonstrates that our CI+CC method is capable of accurately calculating hyperfine-interaction properties, and also implies that the CI+CC method holds promise for precise calculations of other short-range properties.

\subsection{Nuclear quadrupole moment and electric quadrupole hyperfine-structure constants}

\begin{table*}[!]

\caption[]{Electric-field gradients \(q\) (in MHz) for the \(^{3\!}F_{2,3,4}\), \(^{3\!}P_{1,2}\), and \(^{1\!}G_{4}\) states of the \(3d^{2}\) configuration in $^{45}\!$Sc~II, together with the nuclear quadrupole moment of \(^{45}\)Sc. The uncertainties given in parentheses.}
\begin{ruledtabular}
 \begin{tabular}{lcccccccccccccccc}
 States & $q_{\rm M_1}$ & $q_{\rm M_2}$ & $q_{\rm M_3}$  & $q_{\rm M_4}$ & $q_{\rm M_5}$ &$q_{\rm Recc.}$&$B_{\rm Expt.}$~\cite{mansour1989laser} &Q\\
\hline

 $^3\!F_2$ &  46.43 &	49.21 &	  49.52 &	48.67  &	49.46  &49.2(1.0)&	$-10.540(85)$ &$-0.2142(43)$\\
 $^3\!F_3$ &  54.63 &	53.73 &	  54.10 &	53.19  &	54.04  &53.7(1.1)& 	$-12.615(40)$&$-0.2348(47)$\\
 $^3\!F_4$ &  79.85 &	72.08 &	  72.60 &	71.36  &	72.51  &72.1(1.4)&	$-16.456(75)$&$-0.2283(46)$\\
 $^3\!P_1$ &  52.32 &	52.48 &	  52.81 &	52.03  &	52.76  &52.5(1.0)&	$-12.297(6)$  &$-0.2344(47)$\\
 $^3\!P_2$ &$-104.3$ & $-104.8$ &	$-105.5$ & $-103.9$  &   $-105.4$ &$-104.8(2.1)$ &	 22.127(23)&$-0.2110(42)$\\
 $^1\!G_4$ & 274.5 &  295.6 &	 297.3 &	293.0 &	297.0 &295.6(5.9)&	$-63.439(40)$ &$-0.2146(43)$\\
           &        &         &         &          &           &&Final result&$-0.222(5)$\\
          &        &         &         &          &           &&Others          &$-0.220(2)$~\cite{Kell2000CPL}\\
          &        &         &         &          &           &&           &$-0.223(3)$~\cite{Dognon2025PCCP}\\
           &        &         &         &          &           &&           &$-0.231(4)$~\cite{Biero1997pra}\\
\end{tabular}\label{tab5}
\end{ruledtabular}
\end{table*}

Before presenting the electric-quadrupole hyperfine-structure parameter $B$, we briefly recall the nuclear quadrupole moment $Q$ of $^{45}\!$Sc. The presently accepted standard value was established by Kell\"{o} et al.~\cite{Kell2000CPL} from the experimental spectra of the diatomic molecules ScF, ScCl, and ScBr, combined with electric-field-gradient (EFG) $q=B/Q$ calculations at the CCSD(T) level within the Douglas--Kroll one-component approximation. This yielded $Q=-0.220(2)$~b. More recently, Dognon and  Pyykk\"{o} incorporated a new measurement on ScN and performed EFG calculations using the CCSD(T) method in the framework of the four-component Dirac--Coulomb hamiltonian~\cite{Dognon2025PCCP}, obtaining $Q=-0.223(2)$~b. These two $Q$ values obtained from molecular data are in excellent agreement. However, the electric-quadrupole hyperfine constants $B$ for the \(^{3\!}F_{2,3,4}\), \(^{1\!}D_{2}\), \(^{3\!}P_{1,2}\), and \(^{1\!}G_{4}\)  states of the \(3d^{2}\)  configuration in the $^{45}\!$Sc~II have been measured with exceptional precision; if complemented by EFGs of comparable accuracy, these data would allow an equally precise determination of $Q$. Earlier, Bieron et al. employed MCDF calculations for the EFGs of these states and, combined with the experimental $B$ values, extracted $Q=-0.231(4)$~b~\cite{Biero1997pra}. This value is approximately 5\% higher than the standard value of $-0.220(2)$~b~\cite{Kell2000CPL}.

We compute the electric field gradient $q=B/Q$ (in MHz) for the $^{3\!}F_{2,3,4}$, $^{3\!}P_{1,2}$, and $^{1\!}G_{4}$ states of the $3d^{2}$ configuration in the $^{45}$Sc~II using the five methods described above. The results are presented in Table~\ref{tab5}. As with the HFS constants $A$, the CI+LCCSD result is adopted as the recommended value, and its uncertainty is taken to be the larger of (i) the maximum deviation from this value among the three other CI+CC methods and (ii) 2 \% of the recommended value itself. Combining the measured HFS constants $B$ with our EFG values, we extract $6$ independent values of the nuclear quadrupole moment. The final value of the nuclear quadrupole moment $Q$ and its uncertainty are obtained by weighted averaging over the $6$ determinations:
\begin{align}
\begin{cases}
Q_{\rm final}=\dfrac{\displaystyle\sum_{i}^{n}\dfrac{Q_{i}}{(\Delta Q_{i})^2}}{\displaystyle\sum_{i}^{n}\dfrac{1}{(\Delta Q_{i})^2}},\\[16pt]
\Delta Q_{\rm statistic}=\dfrac{1}{\sqrt{\displaystyle\sum_{i}^{n}\dfrac{1}{(\Delta Q_{i})^2}}},\\[16pt]
\chi^2=\dfrac{1}{n-1}\displaystyle\sum_{i}^{n}\dfrac{(Q_{i}-Q_{\rm final})^2}{(\Delta Q_{i})^2},\\[16pt]
\Delta Q_{\rm final}=\Delta Q_{\rm statistic}\times\sqrt{\chi^2}
\end{cases}
\end{align}
where $Q_{i}$ and $\Delta Q_{i}$ denote, respectively, the nuclear quadrupole moment and its uncertainty extracted from the $i$-th atomic state. The reduced chi-squared $\chi^2$ quantifies the consistency of the input values with the weighted mean. The final recommended value and its corresponding uncertainty are given $Q_{\rm final}$ and $\Delta Q_{\rm final}$, respectively.  Given the high precision of the measured HFS constants $B$, the uncertainties in our extracted $Q$ values originate entirely from the electric field gradients. The final result is $Q =-0.222(5)$~b, which is in excellent agreement with values derived from molecular data~\cite{Kell2000CPL,Dognon2025PCCP}.

\begin{table}[!]

\caption[]{Electric quadrupole hyperfine-structure constants (in MHz) of $^{45}\!$Sc II. The uncertainties given in parentheses.}
\begin{ruledtabular}
 \begin{tabular}{lcccccccccccccccc}
Conf & 	Term & $B_{\rm Present}$ &	$B_{\rm Expt.}$	&$B_{\rm Ther.}$ \\
\hline
3d4s & $^3\!D_1$   & --14.3(3)&--13(3)~\cite{Villemoes1992pra} &11.8~\cite{Villemoes1992pra}\\
     &             &          &--11.7(1)(0)~\cite{dockery2023hyperfine}&\\
     & $^3\!D_2$   & --20.4(5)&--30(12)~\cite{Villemoes1992pra} &--13~\cite{Villemoes1992pra}\\
     &             &          &--32.5(2)(0)~\cite{dockery2023hyperfine}&\\
     & $^3\!D_3$   & --40.6(8)&--63(23)~\cite{Villemoes1992pra} &--35.3~\cite{Villemoes1992pra}\\
     &             &          &--45.4(3)(0)~\cite{dockery2023hyperfine}&\\
3d4s & $^1\!D_2$   & --39.5(8)&--39(11)~\cite{Villemoes1992pra}            &--25.5~\cite{Villemoes1992pra}\\
$3d^2$ & $^3\!F_2$ & --10.9(3)&--10.540(85)~\cite{mansour1989laser}     &--8.9~\cite{Villemoes1992pra}\\
       & $^3\!F_3$ & --11.9(3)&--12.615(40)~\cite{mansour1989laser}    &--28.9~\cite{Villemoes1992pra}\\
       & $^3\!F_4$ & --16.0(4)&--16.456(75)~\cite{mansour1989laser}     &--38.1~\cite{Villemoes1992pra}\\
$3d^2$ & $^1\!D_2$ &    7.1(5)&7(7)~\cite{young1988hyperfine}         &10.4~\cite{Villemoes1992pra}\\
       &           &          &7.818(30)~\cite{mansour1989laser}      &\\
$3d^2$ & $^3\!P_1$ & --11.7(3)&--13(2)~\cite{Villemoes1992pra}  &--11.6~\cite{young1988hyperfine}\\
       &           &          &--12.297(6)~\cite{mansour1989laser}&--16.9~\cite{Villemoes1992pra}\\
       & $^3\!P_2$ & 23.3(5)  &26(3)~\cite{Villemoes1992pra}&23.2~\cite{young1988hyperfine}\\
       &           &          &22.127(23)~\cite{mansour1989laser}&18.2~\cite{Villemoes1992pra}	\\
$3d^2$ & $^1\!G_4$ & --65.6(1.3)&--52(32)~\cite{young1988hyperfine}      &--65.4~\cite{young1988hyperfine}	\\
       &           &           &-63.439(40)~\cite{mansour1989laser}& --99.1~\cite{Villemoes1992pra}\\
3d4p & $^1\!D^{o}_2$ &21.7(8)&18(7)~\cite{young1988hyperfine}            &8.4~\cite{young1988hyperfine} \\
     &             &          &                                           &--10.8~\cite{Villemoes1992pra}\\
3d4p & $^3\!F^{o}_2$ & --54.3(1.1)&--40(14)~\cite{young1988hyperfine}     &--47.9~\cite{young1988hyperfine}	\\
     &             &          & --54.7(2)(0)~\cite{dockery2023hyperfine}&--31.7~\cite{Villemoes1992pra}\\
     & $^3\!F^{o}_3$ & --57.9(1.2)&--70(18)~\cite{young1988hyperfine}  &--52.2~\cite{young1988hyperfine}	\\
       &             &          & --59.2(3)(0)~\cite{dockery2023hyperfine}  & --58.5~\cite{Villemoes1992pra}\\
     & $^3\!F^{o}_4$ & --83.7(1.7)&--84(4)~\cite{mansour1989laser}   &--76.0~\cite{Villemoes1992pra}\\
     &             &          & --84.0(9)(0)~\cite{dockery2023hyperfine}      &\\
3d4p & $^3\!D^{o}_1$ &4.7(3)&3(6)~\cite{young1988hyperfine}          &1.1~\cite{young1988hyperfine}	\\
 &             &          &4.9(1)(0)~\cite{dockery2023hyperfine}&--0.7~\cite{Villemoes1992pra} \\
     & $^3\!D^{o}_2$ &3.5(2)&7(7)~\cite{young1988hyperfine}         &--4.7~\cite{young1988hyperfine}	\\
  &             &          &4.8(3)(0)~\cite{dockery2023hyperfine}&--11.7~\cite{Villemoes1992pra}  \\
& $^3\!D^{o}_3$&12.0(3.1)&24(9)~\cite{Villemoes1992pra},        &--5.5~\cite{young1988hyperfine}	\\
  &             &          &13.6(5)(0)~\cite{dockery2023hyperfine} &--13.1~\cite{Villemoes1992pra}\\
3d4p & $^3\!P^{o}_1$ &12.3(3)&9(7)~\cite{young1988hyperfine}&9.5~\cite{young1988hyperfine}	\\
  &             &          &12(6)~\cite{Villemoes1992pra}&11.4~\cite{Villemoes1992pra}\\
     & $^3\!P^{o}_2$ &--21.2(5)&--20(2)~\cite{mansour1989laser} &--17.0~\cite{young1988hyperfine} \\
   &             &          &--21(4)~\cite{Villemoes1992pra}& --13.6~\cite{Villemoes1992pra}\\
3d4p & $^1\!P^{o}_1$ &--25.9(6) &--69(34)~\cite{young1988hyperfine} &--67.0~\cite{young1988hyperfine}	\\
   &             &          &--65(14)~\cite{Villemoes1992pra}& --86.3~\cite{Villemoes1992pra}\\
\end{tabular}\label{tab6}
\end{ruledtabular}
\end{table}

Table~\ref{tab6} presents our calculated HFS constants $B$ for $21$ states of $^{45}$Sc~II, together with available experimental and theoretical values for comparison~\cite{young1988hyperfine,mansour1989laser,Villemoes1992pra,dockery2023hyperfine}. Consistent with our treatment of the HFS constants $A$, we adopt the CI+LCCSD results as the recommended values and assign uncertainties following the same protocol. For the $3d4s$ configuration, our recommended values deviate by approximately 10\% from the most recent measurement~\cite{dockery2023hyperfine}; for other configurations, our results agree reasonably well with the available experimental data within the assigned uncertainties~\cite{young1988hyperfine,mansour1989laser,Villemoes1992pra,dockery2023hyperfine}. Earlier MCDF calculations differ markedly from experiment~\cite{young1988hyperfine,Villemoes1992pra}, with even incorrect signs obtained for several states. In contrast, our CI+CC methods correctly reproduce the experimental signs and achieve significantly improved accuracy compared to previous theoretical work~\cite{young1988hyperfine,Villemoes1992pra}.

It should be noted that the fine-structure intervals in the $3d4s$ configuration are small, only a few tens of cm$^{-1}$; accurate extraction of the HFS constants $B$ for these levels may therefore require second-order corrections arising from off-diagonal hyperfine interactions. Moreover, the HFS constants $B$  of the $^{45}$Sc~II are nearly an order of magnitude smaller than the corresponding HFS constants $A$, placing more stringent demands on measurement precision. We encourage future high-level theoretical calculations and refined measurements to provide independent verification of these HFS constants $B$.

\section{Summary}\label{conclusions}
We have systematically computed the HFS constants $A$ and $B$ for the states arising from the $3d4s$, $3d^2$, $4s^2$, $4s4p$, $3d4p$, $3d5s$, $3d4d$, and $3d5p$ configurations of the $^{45}$Sc~II ion using the relativistic configuration interaction plus coupled-cluster (CI+CC) hybrid method. The calculated HFS constants $A$ agree with high-precision experimental measurements to within approximately 2\% for most states, representing a significant improvement over earlier theoretical calculations. States with small HFS constants $A$ values exhibit stronger sensitivity to high-order electron correlation effects, indicating that further inclusion of triple and quadruple excitations would be necessary to achieve improved accuracy for these challenging cases. The calculated HFS constants $B$ are generally consistent with available experimental observations. From these results, we derive a nuclear electric quadrupole moment $Q=-0.222(5)$~b. This value is in excellent agreement with the value of $Q=-0.223(2)$~b recently obtained from molecular data.

These accurate HFS constants for Sc~II reported in this work constitute indispensable atomic data for stellar elemental abundance determinations.  The demonstrated predictive accuracy of the CI+CC method for for Sc II in this work not only validates the reliability of this theoretical framework but also establishes a robust and promising foundation for systematically extending such calculations to other astrophysically relevant ions, such as Y II, La II, Ce III, and Th III.

\begin{acknowledgments}
The work was supported by the National Natural Science Foundation of China under Grant No.12174268.
\end{acknowledgments}

\begin{thebibliography}{37}%
\makeatletter
\providecommand \@ifxundefined [1]{%
 \@ifx{#1\undefined}
}%
\providecommand \@ifnum [1]{%
 \ifnum #1\expandafter \@firstoftwo
 \else \expandafter \@secondoftwo
 \fi
}%
\providecommand \@ifx [1]{%
 \ifx #1\expandafter \@firstoftwo
 \else \expandafter \@secondoftwo
 \fi
}%
\providecommand \natexlab [1]{#1}%
\providecommand \enquote  [1]{``#1''}%
\providecommand \bibnamefont  [1]{#1}%
\providecommand \bibfnamefont [1]{#1}%
\providecommand \citenamefont [1]{#1}%
\providecommand \href@noop [0]{\@secondoftwo}%
\providecommand \href [0]{\begingroup \@sanitize@url \@href}%
\providecommand \@href[1]{\@@startlink{#1}\@@href}%
\providecommand \@@href[1]{\endgroup#1\@@endlink}%
\providecommand \@sanitize@url [0]{\catcode `\\12\catcode `\$12\catcode
  `\&12\catcode `\#12\catcode `\^12\catcode `\_12\catcode `\%12\relax}%
\providecommand \@@startlink[1]{}%
\providecommand \@@endlink[0]{}%
\providecommand \url  [0]{\begingroup\@sanitize@url \@url }%
\providecommand \@url [1]{\endgroup\@href {#1}{\urlprefix }}%
\providecommand \urlprefix  [0]{URL }%
\providecommand \Eprint [0]{\href }%
\providecommand \doibase [0]{http://dx.doi.org/}%
\providecommand \selectlanguage [0]{\@gobble}%
\providecommand \bibinfo  [0]{\@secondoftwo}%
\providecommand \bibfield  [0]{\@secondoftwo}%
\providecommand \translation [1]{[#1]}%
\providecommand \BibitemOpen [0]{}%
\providecommand \bibitemStop [0]{}%
\providecommand \bibitemNoStop [0]{.\EOS\space}%
\providecommand \EOS [0]{\spacefactor3000\relax}%
\providecommand \BibitemShut  [1]{\csname bibitem#1\endcsname}%
\let\auto@bib@innerbib\@empty
\bibitem [{\citenamefont {Schwartz}(1957)}]{Schwartz}%
  \BibitemOpen
  \bibfield  {author} {\bibinfo {author} {\bibfnamefont {C.}~\bibnamefont
  {Schwartz}},\ }\href {\doibase 10.1103/PhysRev.105.173} {\bibfield  {journal}
  {\bibinfo  {journal} {Phys. Rev.}\ }\textbf {\bibinfo {volume} {105}},\
  \bibinfo {pages} {173} (\bibinfo {year} {1957})}\BibitemShut {NoStop}%
\bibitem [{\citenamefont {{Stone}}(2005)}]{Stone2005ADNDT}%
  \BibitemOpen
  \bibfield  {author} {\bibinfo {author} {\bibfnamefont {N.~J.}\ \bibnamefont
  {{Stone}}},\ }\href {\doibase 10.1016/j.adt.2005.04.001} {\bibfield
  {journal} {\bibinfo  {journal} {At. Data Nucl. Data Tables}\ }\textbf
  {\bibinfo {volume} {90}},\ \bibinfo {pages} {75} (\bibinfo {year}
  {2005})}\BibitemShut {NoStop}%
\bibitem [{\citenamefont {{Pyykk{\"o}}}(2018)}]{Pekka2018MP}%
  \BibitemOpen
  \bibfield  {author} {\bibinfo {author} {\bibfnamefont {P.}~\bibnamefont
  {{Pyykk{\"o}}}},\ }\href {\doibase 10.1080/00268976.2018.1426131} {\bibfield
  {journal} {\bibinfo  {journal} {Mol. Phys.}\ }\textbf {\bibinfo {volume}
  {116}},\ \bibinfo {pages} {1328} (\bibinfo {year} {2018})}\BibitemShut
  {NoStop}%
\bibitem [{\citenamefont {Li}\ \emph {et~al.}(2021)\citenamefont {Li},
  \citenamefont {Qiao}, \citenamefont {Tang},\ and\ \citenamefont
  {Shi}}]{Li2021pra}%
  \BibitemOpen
  \bibfield  {author} {\bibinfo {author} {\bibfnamefont {F.-C.}\ \bibnamefont
  {Li}}, \bibinfo {author} {\bibfnamefont {H.-X.}\ \bibnamefont {Qiao}},
  \bibinfo {author} {\bibfnamefont {Y.-B.}\ \bibnamefont {Tang}}, \ and\
  \bibinfo {author} {\bibfnamefont {T.-Y.}\ \bibnamefont {Shi}},\ }\href
  {\doibase 10.1103/PhysRevA.104.062808} {\bibfield  {journal} {\bibinfo
  {journal} {Phys. Rev. A}\ }\textbf {\bibinfo {volume} {104}},\ \bibinfo
  {pages} {062808} (\bibinfo {year} {2021})}\BibitemShut {NoStop}%
\bibitem [{\citenamefont {de~Groote}\ \emph {et~al.}(2021)\citenamefont
  {de~Groote}, \citenamefont {Kujanp\"a\"a}, \citenamefont {Koszor\'us},
  \citenamefont {Li},\ and\ \citenamefont {Moore}}]{Groote2021pra}%
  \BibitemOpen
  \bibfield  {author} {\bibinfo {author} {\bibfnamefont {R.~P.}\ \bibnamefont
  {de~Groote}}, \bibinfo {author} {\bibfnamefont {S.}~\bibnamefont
  {Kujanp\"a\"a}}, \bibinfo {author} {\bibfnamefont {A.}~\bibnamefont
  {Koszor\'us}}, \bibinfo {author} {\bibfnamefont {J.~G.}\ \bibnamefont {Li}},
  \ and\ \bibinfo {author} {\bibfnamefont {I.~D.}\ \bibnamefont {Moore}},\
  }\href {\doibase 10.1103/PhysRevA.103.032826} {\bibfield  {journal} {\bibinfo
   {journal} {Phys. Rev. A}\ }\textbf {\bibinfo {volume} {103}},\ \bibinfo
  {pages} {032826} (\bibinfo {year} {2021})}\BibitemShut {NoStop}%
\bibitem [{\citenamefont {{Ginges}}\ and\ \citenamefont
  {{Flambaum}}(2004)}]{ginges2004violations}%
  \BibitemOpen
  \bibfield  {author} {\bibinfo {author} {\bibfnamefont {J.~S.~M.}\
  \bibnamefont {{Ginges}}}\ and\ \bibinfo {author} {\bibfnamefont {V.~V.}\
  \bibnamefont {{Flambaum}}},\ }\href {\doibase 10.1016/j.physrep.2004.03.005}
  {\bibfield  {journal} {\bibinfo  {journal} {Phys. Rep.}\ }\textbf {\bibinfo
  {volume} {397}},\ \bibinfo {pages} {63} (\bibinfo {year} {2004})},\ \Eprint
  {http://arxiv.org/abs/physics/0309054} {arXiv:physics/0309054
  [physics.atom-ph]} \BibitemShut {NoStop}%
\bibitem [{\citenamefont {Safronova}\ \emph {et~al.}(2018)\citenamefont
  {Safronova}, \citenamefont {Budker}, \citenamefont {DeMille}, \citenamefont
  {Kimball}, \citenamefont {Derevianko},\ and\ \citenamefont
  {Clark}}]{safronova2018search}%
  \BibitemOpen
  \bibfield  {author} {\bibinfo {author} {\bibfnamefont {M.~S.}\ \bibnamefont
  {Safronova}}, \bibinfo {author} {\bibfnamefont {D.}~\bibnamefont {Budker}},
  \bibinfo {author} {\bibfnamefont {D.}~\bibnamefont {DeMille}}, \bibinfo
  {author} {\bibfnamefont {D.~F.~J.}\ \bibnamefont {Kimball}}, \bibinfo
  {author} {\bibfnamefont {A.}~\bibnamefont {Derevianko}}, \ and\ \bibinfo
  {author} {\bibfnamefont {C.~W.}\ \bibnamefont {Clark}},\ }\href {\doibase
  10.1103/RevModPhys.90.025008} {\bibfield  {journal} {\bibinfo  {journal}
  {Rev. Mod. Phys.}\ }\textbf {\bibinfo {volume} {90}},\ \bibinfo {pages}
  {025008} (\bibinfo {year} {2018})}\BibitemShut {NoStop}%
\bibitem [{\citenamefont {{Del Papa}}\ \emph {et~al.}(2017)\citenamefont {{Del
  Papa}}, \citenamefont {{Holt}},\ and\ \citenamefont
  {{Rosner}}}]{del2017hyperfine}%
  \BibitemOpen
  \bibfield  {author} {\bibinfo {author} {\bibfnamefont {D.~F.}\ \bibnamefont
  {{Del Papa}}}, \bibinfo {author} {\bibfnamefont {R.~A.}\ \bibnamefont
  {{Holt}}}, \ and\ \bibinfo {author} {\bibfnamefont {S.~D.}\ \bibnamefont
  {{Rosner}}},\ }\href {\doibase 10.3390/atoms5010005} {\bibfield  {journal}
  {\bibinfo  {journal} {Atoms}\ }\textbf {\bibinfo {volume} {5}},\ \bibinfo
  {eid} {5} (\bibinfo {year} {2017})}\BibitemShut {NoStop}%
\bibitem [{\citenamefont {{Pickering}}(1996)}]{pickering1996measurements}%
  \BibitemOpen
  \bibfield  {author} {\bibinfo {author} {\bibfnamefont {J.~C.}\ \bibnamefont
  {{Pickering}}},\ }\href {\doibase 10.1086/192382} {\bibfield  {journal}
  {\bibinfo  {journal} {Astrophys. J. Suppl. Ser.}\ }\textbf {\bibinfo {volume}
  {107}},\ \bibinfo {pages} {811} (\bibinfo {year} {1996})}\BibitemShut
  {NoStop}%
\bibitem [{\citenamefont {{Xu}}\ \emph {et~al.}(2021)\citenamefont {{Xu}},
  \citenamefont {{Fang}}, \citenamefont {{Fu}}, \citenamefont {{Liu}},
  \citenamefont {{Ma}}, \citenamefont {{Yang}}, \citenamefont {{Yang}},\ and\
  \citenamefont {{Dai}}}]{xu2021studies}%
  \BibitemOpen
  \bibfield  {author} {\bibinfo {author} {\bibfnamefont {Y.}~\bibnamefont
  {{Xu}}}, \bibinfo {author} {\bibfnamefont {D.}~\bibnamefont {{Fang}}},
  \bibinfo {author} {\bibfnamefont {H.}~\bibnamefont {{Fu}}}, \bibinfo {author}
  {\bibfnamefont {M.}~\bibnamefont {{Liu}}}, \bibinfo {author} {\bibfnamefont
  {H.}~\bibnamefont {{Ma}}}, \bibinfo {author} {\bibfnamefont {Z.}~\bibnamefont
  {{Yang}}}, \bibinfo {author} {\bibfnamefont {Y.}~\bibnamefont {{Yang}}}, \
  and\ \bibinfo {author} {\bibfnamefont {Z.}~\bibnamefont {{Dai}}},\ }\href
  {\doibase 10.1140/epjd/s10053-021-00288-0} {\bibfield  {journal} {\bibinfo
  {journal} {Eur. Phys. J. D}\ }\textbf {\bibinfo {volume} {75}},\ \bibinfo
  {eid} {284} (\bibinfo {year} {2021})}\BibitemShut {NoStop}%
\bibitem [{\citenamefont {{Hala}}\ and\ \citenamefont
  {{Nave}}(2022)}]{nave2022hyperfine}%
  \BibitemOpen
  \bibfield  {author} {\bibinfo {author} {\bibnamefont {{Hala}}}\ and\ \bibinfo
  {author} {\bibfnamefont {G.}~\bibnamefont {{Nave}}},\ }\href {\doibase
  10.3847/1538-4365/ac3edc} {\bibfield  {journal} {\bibinfo  {journal}
  {Astrophys. J. Suppl. Ser.}\ }\textbf {\bibinfo {volume} {259}},\ \bibinfo
  {eid} {17} (\bibinfo {year} {2022})},\ \Eprint
  {http://arxiv.org/abs/2209.11265} {arXiv:2209.11265 [physics.atom-ph]}
  \BibitemShut {NoStop}%
\bibitem [{\citenamefont {{Thorsbro}}\ \emph {et~al.}(2018)\citenamefont
  {{Thorsbro}}, \citenamefont {{Ryde}}, \citenamefont {{Schultheis}},
  \citenamefont {{Hartman}}, \citenamefont {{Rich}}, \citenamefont {{Lomaeva}},
  \citenamefont {{Origlia}},\ and\ \citenamefont
  {{J{\"o}nsson}}}]{thorsbro2018evidence}%
  \BibitemOpen
  \bibfield  {author} {\bibinfo {author} {\bibfnamefont {B.}~\bibnamefont
  {{Thorsbro}}}, \bibinfo {author} {\bibfnamefont {N.}~\bibnamefont {{Ryde}}},
  \bibinfo {author} {\bibfnamefont {M.}~\bibnamefont {{Schultheis}}}, \bibinfo
  {author} {\bibfnamefont {H.}~\bibnamefont {{Hartman}}}, \bibinfo {author}
  {\bibfnamefont {R.~M.}\ \bibnamefont {{Rich}}}, \bibinfo {author}
  {\bibfnamefont {M.}~\bibnamefont {{Lomaeva}}}, \bibinfo {author}
  {\bibfnamefont {L.}~\bibnamefont {{Origlia}}}, \ and\ \bibinfo {author}
  {\bibfnamefont {H.}~\bibnamefont {{J{\"o}nsson}}},\ }\href {\doibase
  10.3847/1538-4357/aadb97} {\bibfield  {journal} {\bibinfo  {journal}
  {Astrophys. J.}\ }\textbf {\bibinfo {volume} {866}},\ \bibinfo {eid} {52}
  (\bibinfo {year} {2018})},\ \Eprint {http://arxiv.org/abs/1808.07489}
  {arXiv:1808.07489 [astro-ph.GA]} \BibitemShut {NoStop}%
\bibitem [{\citenamefont {{Lawler}}\ \emph {et~al.}(2019)\citenamefont
  {{Lawler}}, \citenamefont {{Hala}}, \citenamefont {{Sneden}}, \citenamefont
  {{Nave}}, \citenamefont {{Wood}},\ and\ \citenamefont
  {{Cowan}}}]{lawler2019transition}%
  \BibitemOpen
  \bibfield  {author} {\bibinfo {author} {\bibfnamefont {J.~E.}\ \bibnamefont
  {{Lawler}}}, \bibinfo {author} {\bibnamefont {{Hala}}}, \bibinfo {author}
  {\bibfnamefont {C.}~\bibnamefont {{Sneden}}}, \bibinfo {author}
  {\bibfnamefont {G.}~\bibnamefont {{Nave}}}, \bibinfo {author} {\bibfnamefont
  {M.~P.}\ \bibnamefont {{Wood}}}, \ and\ \bibinfo {author} {\bibfnamefont
  {J.~J.}\ \bibnamefont {{Cowan}}},\ }\href {\doibase 10.3847/1538-4365/ab08ef}
  {\bibfield  {journal} {\bibinfo  {journal} {Astrophys. J. Suppl. Ser.}\
  }\textbf {\bibinfo {volume} {241}},\ \bibinfo {eid} {21} (\bibinfo {year}
  {2019})}\BibitemShut {NoStop}%
\bibitem [{\citenamefont {{Arnesen}}\ \emph {et~al.}(1982)\citenamefont
  {{Arnesen}}, \citenamefont {{Hallin}}, \citenamefont {{Nordling}},
  \citenamefont {{Staaf}}, \citenamefont {{Ward}}, \citenamefont
  {{Jelenkovic}}, \citenamefont {{Kisielinski}}, \citenamefont {{Lundin}},\
  and\ \citenamefont {{Mannervik}}}]{arnesen1982hyperfine}%
  \BibitemOpen
  \bibfield  {author} {\bibinfo {author} {\bibfnamefont {A.}~\bibnamefont
  {{Arnesen}}}, \bibinfo {author} {\bibfnamefont {R.}~\bibnamefont {{Hallin}}},
  \bibinfo {author} {\bibfnamefont {C.}~\bibnamefont {{Nordling}}}, \bibinfo
  {author} {\bibfnamefont {O.}~\bibnamefont {{Staaf}}}, \bibinfo {author}
  {\bibfnamefont {L.}~\bibnamefont {{Ward}}}, \bibinfo {author} {\bibfnamefont
  {B.}~\bibnamefont {{Jelenkovic}}}, \bibinfo {author} {\bibfnamefont
  {M.}~\bibnamefont {{Kisielinski}}}, \bibinfo {author} {\bibfnamefont
  {L.}~\bibnamefont {{Lundin}}}, \ and\ \bibinfo {author} {\bibfnamefont
  {S.}~\bibnamefont {{Mannervik}}},\ }\href@noop {} {\bibfield  {journal}
  {\bibinfo  {journal} {Astron. Astrophys.}\ }\textbf {\bibinfo {volume}
  {106}},\ \bibinfo {pages} {327} (\bibinfo {year} {1982})}\BibitemShut
  {NoStop}%
\bibitem [{\citenamefont {{Mansour}}\ \emph {et~al.}(1989)\citenamefont
  {{Mansour}}, \citenamefont {{Dinneen}}, \citenamefont {{Young}},\ and\
  \citenamefont {{Cheng}}}]{mansour1989laser}%
  \BibitemOpen
  \bibfield  {author} {\bibinfo {author} {\bibfnamefont {N.~B.}\ \bibnamefont
  {{Mansour}}}, \bibinfo {author} {\bibfnamefont {T.}~\bibnamefont
  {{Dinneen}}}, \bibinfo {author} {\bibfnamefont {L.}~\bibnamefont {{Young}}},
  \ and\ \bibinfo {author} {\bibfnamefont {K.~T.}\ \bibnamefont {{Cheng}}},\
  }\href {\doibase 10.1103/PhysRevA.39.5762} {\bibfield  {journal} {\bibinfo
  {journal} {Phys. Rev. A}\ }\textbf {\bibinfo {volume} {39}},\ \bibinfo
  {pages} {5762} (\bibinfo {year} {1989})}\BibitemShut {NoStop}%
\bibitem [{\citenamefont {Young}\ \emph {et~al.}(1988)\citenamefont {Young},
  \citenamefont {Childs}, \citenamefont {Dinneen}, \citenamefont {Kurtz},
  \citenamefont {Berry}, \citenamefont {Engstr\"om},\ and\ \citenamefont
  {Cheng}}]{young1988hyperfine}%
  \BibitemOpen
  \bibfield  {author} {\bibinfo {author} {\bibfnamefont {L.}~\bibnamefont
  {Young}}, \bibinfo {author} {\bibfnamefont {W.~J.}\ \bibnamefont {Childs}},
  \bibinfo {author} {\bibfnamefont {T.}~\bibnamefont {Dinneen}}, \bibinfo
  {author} {\bibfnamefont {C.}~\bibnamefont {Kurtz}}, \bibinfo {author}
  {\bibfnamefont {H.~G.}\ \bibnamefont {Berry}}, \bibinfo {author}
  {\bibfnamefont {L.}~\bibnamefont {Engstr\"om}}, \ and\ \bibinfo {author}
  {\bibfnamefont {K.~T.}\ \bibnamefont {Cheng}},\ }\href {\doibase
  10.1103/PhysRevA.37.4213} {\bibfield  {journal} {\bibinfo  {journal} {Phys.
  Rev. A}\ }\textbf {\bibinfo {volume} {37}},\ \bibinfo {pages} {4213}
  (\bibinfo {year} {1988})}\BibitemShut {NoStop}%
\bibitem [{\citenamefont {Villemoes}\ \emph {et~al.}(1992)\citenamefont
  {Villemoes}, \citenamefont {van Leeuwen}, \citenamefont {Arnesen},
  \citenamefont {Heijkenskj\"old}, \citenamefont {Kastberg}, \citenamefont
  {Larsson},\ and\ \citenamefont {Kotochigova}}]{Villemoes1992pra}%
  \BibitemOpen
  \bibfield  {author} {\bibinfo {author} {\bibfnamefont {P.}~\bibnamefont
  {Villemoes}}, \bibinfo {author} {\bibfnamefont {R.}~\bibnamefont {van
  Leeuwen}}, \bibinfo {author} {\bibfnamefont {A.}~\bibnamefont {Arnesen}},
  \bibinfo {author} {\bibfnamefont {F.}~\bibnamefont {Heijkenskj\"old}},
  \bibinfo {author} {\bibfnamefont {A.}~\bibnamefont {Kastberg}}, \bibinfo
  {author} {\bibfnamefont {M.~O.}\ \bibnamefont {Larsson}}, \ and\ \bibinfo
  {author} {\bibfnamefont {S.~A.}\ \bibnamefont {Kotochigova}},\ }\href
  {\doibase 10.1103/PhysRevA.45.6241} {\bibfield  {journal} {\bibinfo
  {journal} {Phys. Rev. A}\ }\textbf {\bibinfo {volume} {45}},\ \bibinfo
  {pages} {6241} (\bibinfo {year} {1992})}\BibitemShut {NoStop}%
\bibitem [{\citenamefont {{Biero{\'n}}}\ \emph
  {et~al.}(1997{\natexlab{a}})\citenamefont {{Biero{\'n}}}, \citenamefont
  {{Grant}},\ and\ \citenamefont {{Fischer}}}]{bieron1997nuclear}%
  \BibitemOpen
  \bibfield  {author} {\bibinfo {author} {\bibfnamefont {J.}~\bibnamefont
  {{Biero{\'n}}}}, \bibinfo {author} {\bibfnamefont {I.~P.}\ \bibnamefont
  {{Grant}}}, \ and\ \bibinfo {author} {\bibfnamefont {C.~F.}\ \bibnamefont
  {{Fischer}}},\ }\href {\doibase 10.1103/PhysRevA.56.316} {\bibfield
  {journal} {\bibinfo  {journal} {Phys. Rev. A}\ }\textbf {\bibinfo {volume}
  {56}},\ \bibinfo {pages} {316} (\bibinfo {year}
  {1997}{\natexlab{a}})}\BibitemShut {NoStop}%
\bibitem [{\citenamefont {{Biero{\'n}}}\ \emph {et~al.}(1995)\citenamefont
  {{Biero{\'n}}}, \citenamefont {{Parpia}}, \citenamefont {{Fischer}},\ and\
  \citenamefont {{J{\"o}nsson}}}]{bieron1995large}%
  \BibitemOpen
  \bibfield  {author} {\bibinfo {author} {\bibfnamefont {J.}~\bibnamefont
  {{Biero{\'n}}}}, \bibinfo {author} {\bibfnamefont {F.~A.}\ \bibnamefont
  {{Parpia}}}, \bibinfo {author} {\bibfnamefont {C.~F.}\ \bibnamefont
  {{Fischer}}}, \ and\ \bibinfo {author} {\bibfnamefont {P.}~\bibnamefont
  {{J{\"o}nsson}}},\ }\href {\doibase 10.1103/PhysRevA.51.4603} {\bibfield
  {journal} {\bibinfo  {journal} {Phys. Rev. A}\ }\textbf {\bibinfo {volume}
  {51}},\ \bibinfo {pages} {4603} (\bibinfo {year} {1995})}\BibitemShut
  {NoStop}%
\bibitem [{\citenamefont {{Chen}}(1994)}]{chen1994core}%
  \BibitemOpen
  \bibfield  {author} {\bibinfo {author} {\bibfnamefont {G.-x.}\ \bibnamefont
  {{Chen}}},\ }\href {\doibase 10.1016/0375-9601(94)90538-X} {\bibfield
  {journal} {\bibinfo  {journal} {Phys. Let. A}\ }\textbf {\bibinfo {volume}
  {193}},\ \bibinfo {pages} {451} (\bibinfo {year} {1994})}\BibitemShut
  {NoStop}%
\bibitem [{\citenamefont {{Dockery}}\ \emph {et~al.}(2023)\citenamefont
  {{Dockery}}, \citenamefont {{K{\"o}nig}}, \citenamefont {{Lantis}},
  \citenamefont {{Liu}}, \citenamefont {{Minamisono}}, \citenamefont
  {{Pineda}},\ and\ \citenamefont {{Powel}}}]{dockery2023hyperfine}%
  \BibitemOpen
  \bibfield  {author} {\bibinfo {author} {\bibfnamefont {A.}~\bibnamefont
  {{Dockery}}}, \bibinfo {author} {\bibfnamefont {K.}~\bibnamefont
  {{K{\"o}nig}}}, \bibinfo {author} {\bibfnamefont {J.}~\bibnamefont
  {{Lantis}}}, \bibinfo {author} {\bibfnamefont {Y.}~\bibnamefont {{Liu}}},
  \bibinfo {author} {\bibfnamefont {K.}~\bibnamefont {{Minamisono}}}, \bibinfo
  {author} {\bibfnamefont {S.}~\bibnamefont {{Pineda}}}, \ and\ \bibinfo
  {author} {\bibfnamefont {R.}~\bibnamefont {{Powel}}},\ }\href {\doibase
  10.1103/PhysRevA.108.052816} {\bibfield  {journal} {\bibinfo  {journal}
  {Phys. Rev. A}\ }\textbf {\bibinfo {volume} {108}},\ \bibinfo {eid} {052816}
  (\bibinfo {year} {2023})}\BibitemShut {NoStop}%
\bibitem [{\citenamefont {Tang}(2025)}]{Tang2025pra}%
  \BibitemOpen
  \bibfield  {author} {\bibinfo {author} {\bibfnamefont {Y.-B.}\ \bibnamefont
  {Tang}},\ }\href {\doibase 10.1103/l6y6-cf78} {\bibfield  {journal} {\bibinfo
   {journal} {Phys. Rev. A}\ }\textbf {\bibinfo {volume} {112}},\ \bibinfo
  {pages} {042814} (\bibinfo {year} {2025})}\BibitemShut {NoStop}%
\bibitem [{\citenamefont {Tang}(2026)}]{Tang2026pra}%
  \BibitemOpen
  \bibfield  {author} {\bibinfo {author} {\bibfnamefont {Y.-B.}\ \bibnamefont
  {Tang}},\ }\href {\doibase 10.1103/1jxh-c6hw} {\bibfield  {journal} {\bibinfo
   {journal} {Phys. Rev. A}\ }\textbf {\bibinfo {volume} {113}},\ \bibinfo
  {pages} {022812} (\bibinfo {year} {2026})}\BibitemShut {NoStop}%
\bibitem [{\citenamefont {Dzuba}\ \emph {et~al.}(1996)\citenamefont {Dzuba},
  \citenamefont {Flambaum},\ and\ \citenamefont {Kozlov}}]{Dzuba1996pra}%
  \BibitemOpen
  \bibfield  {author} {\bibinfo {author} {\bibfnamefont {V.~A.}\ \bibnamefont
  {Dzuba}}, \bibinfo {author} {\bibfnamefont {V.~V.}\ \bibnamefont {Flambaum}},
  \ and\ \bibinfo {author} {\bibfnamefont {M.~G.}\ \bibnamefont {Kozlov}},\
  }\href {\doibase 10.1103/PhysRevA.54.3948} {\bibfield  {journal} {\bibinfo
  {journal} {Phys. Rev. A}\ }\textbf {\bibinfo {volume} {54}},\ \bibinfo
  {pages} {3948} (\bibinfo {year} {1996})}\BibitemShut {NoStop}%
\bibitem [{\citenamefont {Tang}\ \emph {et~al.}(2017)\citenamefont {Tang},
  \citenamefont {Lou},\ and\ \citenamefont {Shi}}]{Tang2017pra}%
  \BibitemOpen
  \bibfield  {author} {\bibinfo {author} {\bibfnamefont {Y.-B.}\ \bibnamefont
  {Tang}}, \bibinfo {author} {\bibfnamefont {B.-Q.}\ \bibnamefont {Lou}}, \
  and\ \bibinfo {author} {\bibfnamefont {T.-Y.}\ \bibnamefont {Shi}},\ }\href
  {\doibase 10.1103/PhysRevA.96.022513} {\bibfield  {journal} {\bibinfo
  {journal} {Phys. Rev. A}\ }\textbf {\bibinfo {volume} {96}},\ \bibinfo
  {pages} {022513} (\bibinfo {year} {2017})}\BibitemShut {NoStop}%
\bibitem [{\citenamefont {Blundell}\ \emph {et~al.}(1989)\citenamefont
  {Blundell}, \citenamefont {Johnson}, \citenamefont {Liu},\ and\ \citenamefont
  {Sapirstein}}]{Blundell1989pra}%
  \BibitemOpen
  \bibfield  {author} {\bibinfo {author} {\bibfnamefont {S.~A.}\ \bibnamefont
  {Blundell}}, \bibinfo {author} {\bibfnamefont {W.~R.}\ \bibnamefont
  {Johnson}}, \bibinfo {author} {\bibfnamefont {Z.~W.}\ \bibnamefont {Liu}}, \
  and\ \bibinfo {author} {\bibfnamefont {J.}~\bibnamefont {Sapirstein}},\
  }\href {\doibase 10.1103/PhysRevA.40.2233} {\bibfield  {journal} {\bibinfo
  {journal} {Phys. Rev. A}\ }\textbf {\bibinfo {volume} {40}},\ \bibinfo
  {pages} {2233} (\bibinfo {year} {1989})}\BibitemShut {NoStop}%
\bibitem [{\citenamefont {{Dzuba}}\ \emph {et~al.}(1998)\citenamefont
  {{Dzuba}}, \citenamefont {{Flambaum}}, \citenamefont {{Kozlov}},\ and\
  \citenamefont {{Porsev}}}]{Dzuba1998jetp}%
  \BibitemOpen
  \bibfield  {author} {\bibinfo {author} {\bibfnamefont {V.~A.}\ \bibnamefont
  {{Dzuba}}}, \bibinfo {author} {\bibfnamefont {V.~V.}\ \bibnamefont
  {{Flambaum}}}, \bibinfo {author} {\bibfnamefont {M.~G.}\ \bibnamefont
  {{Kozlov}}}, \ and\ \bibinfo {author} {\bibfnamefont {S.~G.}\ \bibnamefont
  {{Porsev}}},\ }\href {\doibase 10.1134/1.558736} {\bibfield  {journal}
  {\bibinfo  {journal} {J. Exp. Theor. Phys.}\ }\textbf {\bibinfo {volume}
  {87}},\ \bibinfo {pages} {885} (\bibinfo {year} {1998})}\BibitemShut
  {NoStop}%
\bibitem [{\citenamefont {{Safronova}}\ \emph {et~al.}(1999)\citenamefont
  {{Safronova}}, \citenamefont {{Derevianko}}, \citenamefont {{Safronova}},\
  and\ \citenamefont {{Johnson}}}]{Safronova1999jpb}%
  \BibitemOpen
  \bibfield  {author} {\bibinfo {author} {\bibfnamefont {U.~I.}\ \bibnamefont
  {{Safronova}}}, \bibinfo {author} {\bibfnamefont {A.}~\bibnamefont
  {{Derevianko}}}, \bibinfo {author} {\bibfnamefont {M.~S.}\ \bibnamefont
  {{Safronova}}}, \ and\ \bibinfo {author} {\bibfnamefont {W.~R.}\ \bibnamefont
  {{Johnson}}},\ }\href {\doibase 10.1088/0953-4075/32/14/319} {\bibfield
  {journal} {\bibinfo  {journal} {J. Phys. B: At. Mol. Phys.}\ }\textbf
  {\bibinfo {volume} {32}},\ \bibinfo {pages} {3527} (\bibinfo {year}
  {1999})}\BibitemShut {NoStop}%
\bibitem [{\citenamefont {Savukov}(2004)}]{Savukov2004pra}%
  \BibitemOpen
  \bibfield  {author} {\bibinfo {author} {\bibfnamefont {I.~M.}\ \bibnamefont
  {Savukov}},\ }\href {\doibase 10.1103/PhysRevA.70.042502} {\bibfield
  {journal} {\bibinfo  {journal} {Phys. Rev. A}\ }\textbf {\bibinfo {volume}
  {70}},\ \bibinfo {pages} {042502} (\bibinfo {year} {2004})}\BibitemShut
  {NoStop}%
\bibitem [{\citenamefont {Safronova}\ \emph {et~al.}(2009)\citenamefont
  {Safronova}, \citenamefont {Kozlov}, \citenamefont {Johnson},\ and\
  \citenamefont {Jiang}}]{Safronova2009pra}%
  \BibitemOpen
  \bibfield  {author} {\bibinfo {author} {\bibfnamefont {M.~S.}\ \bibnamefont
  {Safronova}}, \bibinfo {author} {\bibfnamefont {M.~G.}\ \bibnamefont
  {Kozlov}}, \bibinfo {author} {\bibfnamefont {W.~R.}\ \bibnamefont {Johnson}},
  \ and\ \bibinfo {author} {\bibfnamefont {D.}~\bibnamefont {Jiang}},\ }\href
  {\doibase 10.1103/PhysRevA.80.012516} {\bibfield  {journal} {\bibinfo
  {journal} {Phys. Rev. A}\ }\textbf {\bibinfo {volume} {80}},\ \bibinfo
  {pages} {012516} (\bibinfo {year} {2009})}\BibitemShut {NoStop}%
\bibitem [{\citenamefont {Dzuba}(2014)}]{Dzuba2014pra}%
  \BibitemOpen
  \bibfield  {author} {\bibinfo {author} {\bibfnamefont {V.~A.}\ \bibnamefont
  {Dzuba}},\ }\href {\doibase 10.1103/PhysRevA.90.012517} {\bibfield  {journal}
  {\bibinfo  {journal} {Phys. Rev. A}\ }\textbf {\bibinfo {volume} {90}},\
  \bibinfo {pages} {012517} (\bibinfo {year} {2014})}\BibitemShut {NoStop}%
\bibitem [{\citenamefont {Tang}\ \emph {et~al.}(2013)\citenamefont {Tang},
  \citenamefont {Qiao}, \citenamefont {Shi},\ and\ \citenamefont
  {Mitroy}}]{Tang2013pra}%
  \BibitemOpen
  \bibfield  {author} {\bibinfo {author} {\bibfnamefont {Y.-B.}\ \bibnamefont
  {Tang}}, \bibinfo {author} {\bibfnamefont {H.-X.}\ \bibnamefont {Qiao}},
  \bibinfo {author} {\bibfnamefont {T.-Y.}\ \bibnamefont {Shi}}, \ and\
  \bibinfo {author} {\bibfnamefont {J.}~\bibnamefont {Mitroy}},\ }\href
  {\doibase 10.1103/PhysRevA.87.042517} {\bibfield  {journal} {\bibinfo
  {journal} {Phys. Rev. A}\ }\textbf {\bibinfo {volume} {87}},\ \bibinfo
  {pages} {042517} (\bibinfo {year} {2013})}\BibitemShut {NoStop}%
\bibitem [{\citenamefont {Kramida}\ \emph {et~al.}(2024)\citenamefont
  {Kramida}, \citenamefont {{Yu.~Ralchenko}}, \citenamefont {Reader},\ and\
  \citenamefont {{and NIST ASD Team}}}]{NIST_ASD}%
  \BibitemOpen
  \bibfield  {author} {\bibinfo {author} {\bibfnamefont {A.}~\bibnamefont
  {Kramida}}, \bibinfo {author} {\bibnamefont {{Yu.~Ralchenko}}}, \bibinfo
  {author} {\bibfnamefont {J.}~\bibnamefont {Reader}}, \ and\ \bibinfo {author}
  {\bibnamefont {{and NIST ASD Team}}},\ }\href@noop {} {}\bibinfo
  {howpublished} {{NIST Atomic Spectra Database (ver. 5.12), [Online].
  Available: {doi:https://doi.org/10.18434/T4W30F}. National Institute of
  Standards and Technology, Gaithersburg, MD.}} (\bibinfo {year}
  {2024})\BibitemShut {NoStop}%
\bibitem [{\citenamefont {{Ruczkowski}}\ \emph {et~al.}(2014)\citenamefont
  {{Ruczkowski}}, \citenamefont {{Elantkowska}},\ and\ \citenamefont
  {{Dembczy{\'n}ski}}}]{Ruczkowski2014jsqrt}%
  \BibitemOpen
  \bibfield  {author} {\bibinfo {author} {\bibfnamefont {J.}~\bibnamefont
  {{Ruczkowski}}}, \bibinfo {author} {\bibfnamefont {M.}~\bibnamefont
  {{Elantkowska}}}, \ and\ \bibinfo {author} {\bibfnamefont {J.}~\bibnamefont
  {{Dembczy{\'n}ski}}},\ }\href {\doibase 10.1016/j.jqsrt.2014.04.018}
  {\bibfield  {journal} {\bibinfo  {journal} {J. Quant. Spectrosc. Radiat.
  Transf.}\ }\textbf {\bibinfo {volume} {145}},\ \bibinfo {pages} {20}
  (\bibinfo {year} {2014})}\BibitemShut {NoStop}%
\bibitem [{\citenamefont {{Kell{\"o}}}\ \emph {et~al.}(2000)\citenamefont
  {{Kell{\"o}}}, \citenamefont {{Sadlej}},\ and\ \citenamefont
  {{Pyykk{\"o}}}}]{Kell2000CPL}%
  \BibitemOpen
  \bibfield  {author} {\bibinfo {author} {\bibfnamefont {V.}~\bibnamefont
  {{Kell{\"o}}}}, \bibinfo {author} {\bibfnamefont {A.~J.}\ \bibnamefont
  {{Sadlej}}}, \ and\ \bibinfo {author} {\bibfnamefont {P.}~\bibnamefont
  {{Pyykk{\"o}}}},\ }\href {\doibase 10.1016/S0009-2614(00)00946-5} {\bibfield
  {journal} {\bibinfo  {journal} {Chem. Phys. Lett.}\ }\textbf {\bibinfo
  {volume} {329}},\ \bibinfo {pages} {112} (\bibinfo {year}
  {2000})}\BibitemShut {NoStop}%
\bibitem [{\citenamefont {{Dognon}}\ and\ \citenamefont
  {{Pyykk{\"o}}}(2025)}]{Dognon2025PCCP}%
  \BibitemOpen
  \bibfield  {author} {\bibinfo {author} {\bibfnamefont {J.-P.}\ \bibnamefont
  {{Dognon}}}\ and\ \bibinfo {author} {\bibfnamefont {P.}~\bibnamefont
  {{Pyykk{\"o}}}},\ }\href {\doibase 10.1039/D5CP01943E} {\bibfield  {journal}
  {\bibinfo  {journal} {Phys. Chem. Chem. Phys.}\ }\textbf {\bibinfo {volume}
  {27}},\ \bibinfo {pages} {20453} (\bibinfo {year} {2025})}\BibitemShut
  {NoStop}%
\bibitem [{\citenamefont {{Biero{\'n}}}\ \emph
  {et~al.}(1997{\natexlab{b}})\citenamefont {{Biero{\'n}}}, \citenamefont
  {{Grant}},\ and\ \citenamefont {{Fischer}}}]{Biero1997pra}%
  \BibitemOpen
  \bibfield  {author} {\bibinfo {author} {\bibfnamefont {J.}~\bibnamefont
  {{Biero{\'n}}}}, \bibinfo {author} {\bibfnamefont {I.~P.}\ \bibnamefont
  {{Grant}}}, \ and\ \bibinfo {author} {\bibfnamefont {C.~F.}\ \bibnamefont
  {{Fischer}}},\ }\href {\doibase 10.1103/PhysRevA.56.316} {\bibfield
  {journal} {\bibinfo  {journal} {\pra}\ }\textbf {\bibinfo {volume} {56}},\
  \bibinfo {pages} {316} (\bibinfo {year} {1997}{\natexlab{b}})}\BibitemShut
  {NoStop}%
\end{thebibliography}
%
\end{document}